\documentclass[manuscript]{acmart}

\AtBeginDocument{%
  }

\usepackage{graphicx}
\usepackage{subfigure}
\usepackage{multirow}
\usepackage{calc}

\usepackage{textcomp}
\usepackage{siunitx}
\usepackage{soul,xcolor}
\newcommand{\quoting}[2][P]{``\emph{#2}''\emph{[\textnormal{#1}]}}

\usepackage{longtable}
\newcommand{\purple}[1]{\textcolor{purple}{#1}}

\copyrightyear{2024}
\acmYear{2024}
\setcopyright{rightsretained}
\acmConference[CHI EA '24]{Extended Abstracts of the CHI Conference on
Human Factors in Computing Systems}{May 11--16, 2024}{Honolulu, HI, USA}
\acmBooktitle{Extended Abstracts of the CHI Conference on Human Factors in
Computing Systems (CHI EA '24), May 11--16, 2024, Honolulu, HI, USA}
\acmDOI{10.1145/3613905.3650750}
\acmISBN{979-8-4007-0331-7/24/05}

\begin{document}



\title[Transparent AI Disclosure Obligations]{Transparent AI Disclosure Obligations:\\ Who, What, When, Where, Why, How}
\author{Abdallah El Ali}
\affiliation{
  \institution{Centrum Wiskunde \& Informatica}
  \city{Amsterdam}
  \country{Netherlands}}
\email{aea@cwi.nl}
\orcid{0000-0002-9954-4088}

\author{Karthikeya Puttur Venkatraj}
\affiliation{
  \institution{Centrum Wiskunde \& Informatica}
  \city{Amsterdam}
  \country{Netherlands}}
\email{karthike@cwi.nl}
\orcid{0009-0003-4245-8802}

\author{Sophie Morosoli}
\affiliation{
  \institution{University of Amsterdam}
  \city{Amsterdam}
  \country{Netherlands}}
\email{s.v.morosoli@uva.nl}
\orcid{0000-0002-4651-3819}

\author{Laurens Naudts}
\affiliation{
 \institution{University of Amsterdam}
  \city{}
  \country{}}
 \affiliation{%
  \institution{KU Leuven}
  \city{Amsterdam}
  \country{Netherlands}}
\email{l.p.a.naudts@uva.nl}
\orcid{0000-0002-5777-1450}

\author{Natali Helberger}
\affiliation{
  \institution{University of Amsterdam}
  \city{Amsterdam}
  \country{Netherlands}}
\email{n.helberger@uva.nl}
\orcid{0000-0003-1652-0580}

\author{Pablo Cesar}
\affiliation{%
  \institution{Centrum Wiskunde \& Informatica}
  \city{}
  \country{}}
  \affiliation{%
  \institution{Delft University of Technology}
  \city{Amsterdam}
  \country{The Netherlands}}
\email{p.s.cesar@cwi.nl}
\orcid{0000-0003-1752-6837}

\renewcommand{\shortauthors}{El Ali et al.}

\begin{abstract}

Advances in Generative Artificial Intelligence (AI) are resulting in AI-generated media output that is (nearly) indistinguishable from human-created content. This can drastically impact users and the media sector, especially given global risks of misinformation. While the currently discussed European AI Act aims at addressing these risks through Article 52's AI transparency obligations, its interpretation and implications remain unclear. In this early work, we adopt a participatory AI approach to derive key questions based on Article 52's disclosure obligations. We ran two workshops with researchers, designers, and engineers across disciplines (N=16), where participants deconstructed Article 52's relevant clauses using the 5W1H framework. We contribute a set of 149 questions clustered into five themes and 18 sub-themes. We believe these can not only help inform future legal developments and interpretations of Article 52, but also provide a starting point for Human-Computer Interaction research to (re-)examine disclosure transparency from a human-centered AI lens.

\end{abstract}





\maketitle

\section{Introduction}

\begin{flushright}
\small{\textit{``Fake realities will create fake humans. Or, fake humans will generate fake realities and then sell them to other humans, turning them, eventually, into forgeries of themselves."}} \\
\small{--- Philip K. Dick, \textit{I Hope I Shall Arrive Soon,} 1980}
\end{flushright}
\noindent
The field of Artificial Intelligence (AI), and the media sector at large, are undergoing a transformative change. Foundation models, developed on the basis of deep neural networks and self-supervised learning, have gained widespread acceptance \cite{Bommasani2021FoundationModels}. These models have led to the emergence of Generative AI (GenAI) tools like Midjourney and ChatGPT, and have shown impressive capabilities in producing media content, including images \cite{saharia2022photorealistic}, text \cite{openai2023}, videos \cite{ho2022imagenvideo}, and audio \cite{borsos2022audiolm}. These have the potential to drastically impact users and the media sector, essentially blurring the line between fiction and reality as users engage with media. Model output has reached a level by which humans can no longer perceive GenAI outputs as distinguishable from human-generated content \cite{groh2023human}. What digital media appears to be true and authentic cannot necessarily be trusted, and is reportedly fueling the spread of mis- and disinformation\footnote{https://www.axios.com/2023/02/21/chatbots-misinformation-nightmare-chatgpt-ai}. Therefore it is imperative to disclose the use of AI in the generation/manipulation of media content. According to the World Economic Forum's 2024 Global Risks Report \cite{WEF2024}, ``misinformation and disinformation" were ranked as the highest global risk anticipated over the next two years. This also comes at a time where OpenAI issues a statement regarding the upcoming 2024 worldwide elections \cite{openaiOpenAIApproaching}, emphasizing: ``Transparency around AI-generated content". In fact, in the European Union (EU), such an obligation will be incorporated into the upcoming AI Act \cite{EUAIAct}. Despite that the AI Act is still in development and subject to change, the core obligation to ``disclose AI-driven interaction" appears to be stable.

Algorithmic approaches to automatically detect text (co-)produced using GenAI tools (e.g., ChatGPT) are so far unreliable \cite{sadasivan2023aigenerated}, and can result in consistent biases against specific user groups (e.g., non-native English writers \cite{Liang2023}). While measures to encode provenance cryptographically in images (cf., Coalition for Content Provenance and Authenticity \cite{c2paIntroducingOfficial}) or audio (SynthID \cite{SynthID}) are being taken, these are not widely implemented. Moreover, it remains unclear how these need to be displayed to users. Despite concerted efforts worldwide, constructing even simple disclosure measures (e.g., watermarks) that actually succeed in reducing public risks is itself a difficult task. Disclosure is defined as ``to make known or public" \cite{DiscloseDefinition}, where in the context of AI, disclosures can ``contain information about the data collection, data processing, and decision-making practices of a digital product and are voluntarily provided by the product's vendor (an individual developer or an organization)." (Open Ethics Initiative \cite{openethicsTaxonomy}). A fundamental challenge here is in ensuring effective transparency measures, given the rapid pace of AI development and deployment. To mitigate harms and risks, regulatory efforts such as the AI Disclosure Act of 2023 in the United States \cite{Torres2023} and the EU AI Act \cite{EUAIAct} (currently under discussion \cite{AIActupdates}) aim to tackle this. In this work, we focus on the EU AI Act proposition (Article 52: ``Transparency obligations for providers and users of certain AI systems"), that addresses the issue of AI system transparency\footnote{Even though the final text of the AI Act must still be agreed upon, the inclusion of a "disclosure obligation" is (almost) a certitude.}. 

In this early work, we adopt a human-centered approach to derive key questions that arise around transparent AI disclosures. We ask: \textbf{RQ:} What are the key considerations and concerns surrounding transparent AI disclosures in the context of the EU AI Act? We ran two participatory AI workshops with researchers, designers, and engineers across disciplines (N=16), where we utilized the 5W1H framework \cite{Hart1996,Jia2016} to deconstruct the relevant clauses in Article 52 concerning AI disclosures for users and providers. This was done in the context of the media sector, with a focus on media consumption and production. Even if the language of Article 52 changes, our goal is to trigger reflection on disclosures and what these might look like. We contribute a set of 149 questions clustered into five themes and 18 sub-themes, that we anticipate can help drive interdisciplinary research forward in responsible AI. Our work aims at tackling the challenge of interpreting and implementing obligatory transparent AI disclosures for ever-evolving AI technology in an interdisciplinary manner. This not only helps inform legal developments and future interpretations of Article 52, but also provides a starting point for the Human-Computer Interaction (HCI) community to (re-)examine disclosures from a human-centered AI lens.

\section{Background and Related Work}

\subsection{EU AI Act's Article 52 and AI disclosure implications}

We focus on the EU AI Act \cite{EUAIAct} (currently under discussion \cite{AIActupdates}) proposition (Article 52: ``Transparency obligations for providers and users of certain AI systems"), that addresses AI system transparency. It raises two important clauses:

\begin{itemize}
\item ``$\S$ 1. Providers shall ensure that Al systems intended to interact with natural persons are designed and developed in such a way that natural persons are informed that they are interacting with an Al system unless this is obvious from the circumstances and the context of use."

\item ``$\S$ 3. Users of an Al system that generates or manipulates image, audio or video content that appreciably resembles existing persons, objects, places or other entities or events and would falsely appear to a person to be authentic or truthful ('deep fake'), shall disclose that the content has been artificially generated or manipulated."
\end{itemize}

From these clauses, the types of disclosure are unspecified, and terms such as `authentic' are undefined, leaving them open for interpretation. For example, `what', `how', or `when' should such AI disclosures take place, even if automatically detected, remains unclear. While the most recent draft of the AI Act \cite{techmonitorTextLeaked} touches on aspects of this, where Art. 52(3) (second subparagraph) highlights when to disclose, much remains uncharted regarding human oversight processes when exceptions are made. We believe it is important to carefully assess the AI Act's disclosure implications, before proceeding to Human-AI interactions, wherein explainability \cite{Panigutti2023} may be an additional challenge. Furthermore, authenticity is itself a multidimensional concept \cite{kernis_multicomponent_2006}, which raises the question of what aspects of inauthentic media need to be disclosed. AI disclosures, be they automatic (algorithmic) or provided by system designers or other stakeholders, may include the following: revealing when content is AI-generated, registering these emerging AI systems with a database, summarizing copyrighted material used in training these systems, publication of risk assessments, or even trust certification labels \cite{scharowski_certification_2023}. Importantly, encounters with AI-generated content can impact the human experience of algorithms, and more broadly the psychology of Human-AI interaction \cite{sundar2020rise}, to which we turn to next. 

\subsection{Human-AI interaction, media consumption, and transparent AI disclosures}

For scholarly work, the Association for Computing Machinery has instated clear policies on GenAI, stating ``The use of generative AI tools and technologies to create content is permitted but must be fully disclosed in the Work." \cite{ACMauthorship}. This contributes to ongoing discussions on the ethics of AI disclosure in scholarly works \cite{Hosseini2023}. Within HCI research and practice, Schmidt et al. \cite{Schmidt2024}, in rethinking Human-Centered Design in the age of GenAI, emphasize the need for being ``transparent and honest" when it comes to AI tool usage. Indeed, issues of ownership and agency arguably span the entire HCI research cycle \cite{Elagroudy2024}. Such interactions will continue to pervade not just scholarly discourse, but also the everyday media we consume, making it crucial to understand the impact on human perceptions. For algorithmic decision making, Langer et al. \cite{Langer2022} showed that terminology (e.g., `algorithms' vs. `artificial intelligence') affects laypeople's perceptions of system properties and evaluations (e.g., trust) -- they recommend being mindful when choosing terms given unintended consequences, and their impact on HCI research robustness and replicability. Within COVID-19 health (mis-)information, Jia et al. \cite{Jia2022} found that various misinformation labels (e.g., algorithm, community, third-party fact-checker) are dependent on people's political ideology (liberal, conservative). Cloudy et al. \cite{cloudy_straight_2023} found that a news story presented as sourced from an AI journalist activated individuals' machine heuristic (rule of thumb that machines are more secure and trustworthy than humans \cite{Sundar2019}), which helps mitigate the hostile media bias effect. Furthermore, there may be hidden dangers in such labeling approaches, where they may lead users to believe that content that is not labeled is actually factual, when it may not be -- the so-called ``implied truth effect" \cite{Pennycook2020}. Epstein et al. \cite{Epstein2023} found that participants consistently associated ``AI Generated," ``Generated with an AI tool," and ``AI manipulated" with AI content out of nine potential labels, regardless of whether or not they were misleading. These works underscore the importance of disclosure-based interventions, and highlight the wild west of today's disclosure approaches, from tool usage to perceiving and understanding disclosure labels to safeguard against fake news.

\section{Method}

\subsection{Workshops: approach and objectives}

To generate key questions in a participatory manner, we ran two workshops in December (2023) and January (2024). We utilized the 5W1H (Who, What, When, Where, Why, How) framework, which is widely used in journalism \cite{Hart1996}. Furthermore, this framework allows defining a high-dimensional design space (cf., software testing for cloud computing \cite{Jia2016}), by which an initial set of research questions can be systematically formulated. Our objective was to help deconstruct the relevant clauses in Article 52 by identifying key questions related to the interpretation, implementation, and societal impact of AI disclosures for users and providers. This was done in the context of the media sector, with a focus on media consumption and production. We anticipated that this would help unravel core values, consideration, and risks behind each question type. Workshops were conducted at two different locations/institutes, each targeting a specific researcher demographic. At the first location (session 1), participants had expertise from law, political science, communication science, and artificial intelligence. At the second location (session 2), we had expertise from computer science, human-computer interaction, and (interaction) design. For each workshop session, we collected: demographics, informed consent, photos of the session, and the resulting questions generated.

\subsection{Workshop materials}

\begin{figure*}[t]
	\centering
	\subfigure[Workshop Session 1]{\label{fig:rsp}\includegraphics[width=0.38\linewidth]{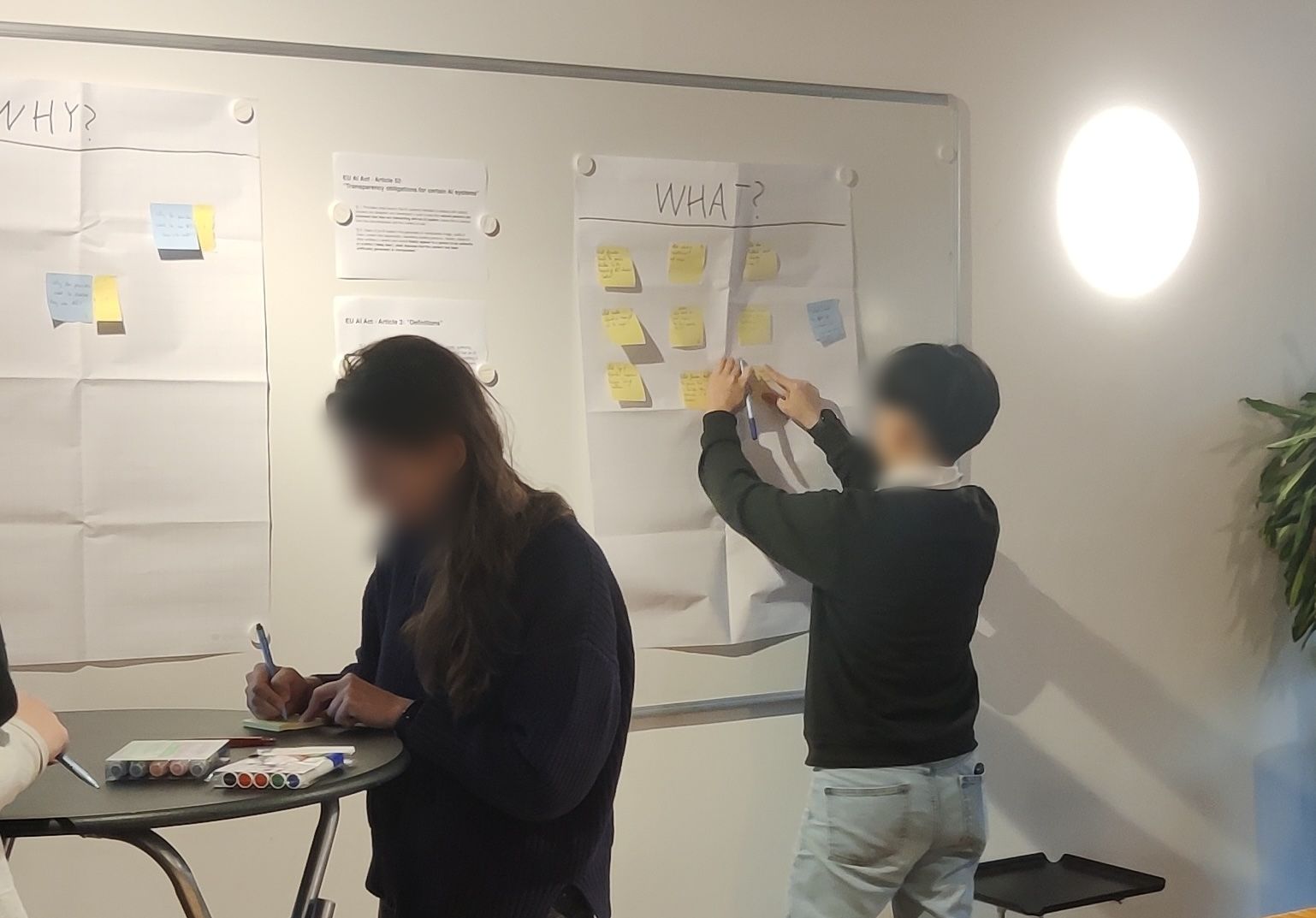}}
	\subfigure[Workshop Session 2]{\label{fig:tasks}\includegraphics[width=0.43\linewidth]{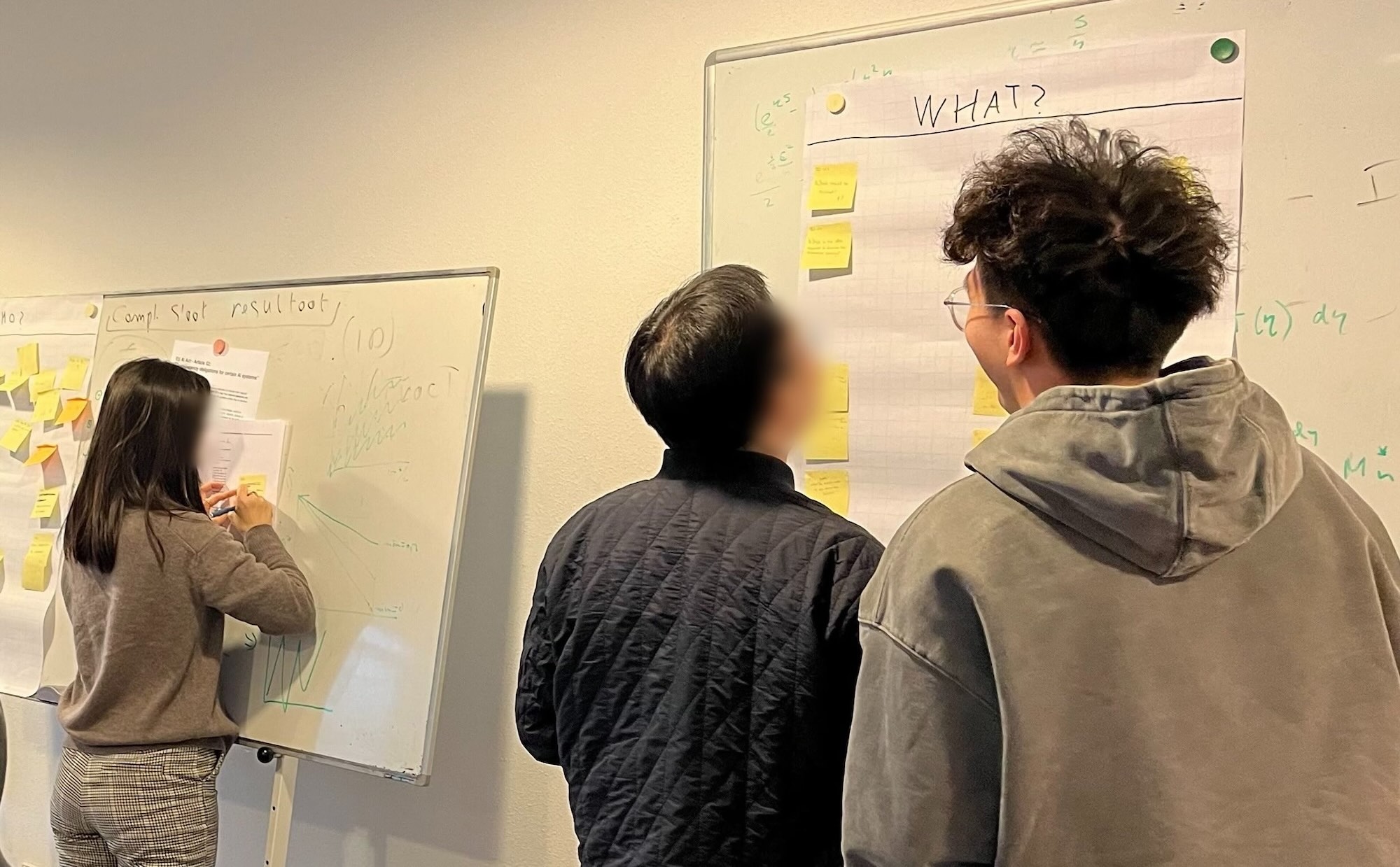}}
	\caption{Snapshots of the workshop sessions showing participants generating questions.}
	\Description{Two images side by side showing participants in our workshops. The left image shows the workshop that took place at the first location targeting mainly those with a legal and policy background, and the second shows the workshop that look place at the second location, targeting mainly those with a computing and design background. In both photos, participants are writing on sticky notes on a flip-chart of a question type.}
	\label{fig:sessions}
\end{figure*}

The workshop setup consisted of six flip charts that were attached to the walls of the workshop area (see Figure \ref{fig:sessions}). Each flip chart was given a heading of one of the 5W1H terms to indicate to participants which type of questions to ask there. Post-its and markers were provided. Article 52 and Article 3 were printed on an A4 landscape paper, and placed near each flip chart for quick reference. Article 3 provides definitions of `provider' and `user'\footnote{In the latest draft version of the EU AI Act (leaked online on January 21, 2024) \cite{techmonitorTextLeaked}, `user' is now referred to as `deployer'. However, this does not impact our workshop outcomes, even if participants interpreted `user' as intended by the law (`deployer`) or as `recipient'.}, that helped participants have a clear understanding of what these `legal' terms mean. Article 3 (current revision; bold added for emphasis) states:

\begin{itemize}
\item ``$\S$ 1. \textbf{`provider'} means a natural or legal person, public authority, agency or other body that \textbf{develops an AI system} or that has an AI system developed with a view to placing it on the market or putting it into service under its own name or trademark, whether for payment or free of charge?" 
\item ``$\S$ 4. \textbf{`user'} means any natural or legal person, public authority, agency or other body \textbf{using an AI system} under its authority, except where the AI system is used in the course of a personal non-professional activity."
\end{itemize}

\subsection{Study procedure}

Sessions began with a short ~10 minute introduction to set the workshop context, to obtain participant consent (data protection and privacy), and for them to fill in a demographics form. Each workshop session consisted of three tasks, each lasting around 25-30 min., amounting to approximately one and a half hours in total. In the introduction, we first presented the EU AI Act, along with Article 52 and a brief explanation of disclosures. Participants were assigned an ID to note down with each post-it note. Participants were split into groups of two or three, depending on the total size in the session. The grouping was only for discussion and logistical purposes, so each question posed was still done individually. Task 1 and 2 were question generation tasks, where participants were instructed to individually generate questions for each of the 5W1H questions. After 10 minutes, they moved to a different question, such that at the end of ~60 minutes each group has had the chance to provide their input for all six question types. Participants were also instructed to review the questions that were placed by participants on the flip chart, and to optionally vote with a +/- 1 post-it on the question if they were in agreement/disagreement. For Task 3, we divided each flip chart into two parts: users and providers. This provided participants the opportunity to reflect on and consider whether the posted questions were relevant for the user, the provider, or both (in which case it was placed in the center of the flip chart). Participants were offered candy in the session. When the session ended, participants were thanked for their time and efforts.

\subsection{Participants}

We had 16 participants (8 female, 7 male, 1 non-binary), where the first workshop had seven, and the second nine participants. Twelve were in the 25-34 age group, three in the 18-24 age group, and one in the 35-44 age group. Three were pursuing their master's degree, seven either their PhD or had research assistant roles, and six had completed their PhD with researcher roles (postdoctoral or higher). Participants' affiliations were spread across five different institutes. Nine had expertise in computer science and engineering (incl. computer vision, natural language processing, and signal processing), two in HCI and design, two in communication science, two in law, and one in political science.

\subsection{Analysis approach}
We analyzed our data using inductive thematic analysis \cite{clarke2015thematic}. First, the first author created early codes for the 155 questions. These codes were clustered into different topics (e.g., `misinformation', `trust', `ux', `infovis'). Based on these, an initial set of sub-themes and themes emerged. The second author independently reviewed all questions and corresponding sub- and main themes. Instead of calculating statistical inter-rater reliability for the analysis, the consensus among the authors was reached through two online video discussions (each approximately 1.5 hours) \cite{McDonald2019reliability}, where each question was carefully re-assessed by both researchers to arrive at a final list of sub-themes and themes.

\section{Results}

Our workshops resulted in a total of 155 questions, where the raw digitized data can be seen in \purple{Supplementary Material}. After data cleaning (removing near-duplicates, irrelevant questions), this resulted in 149 questions. Questions were edited for typos, but left largely as is. Our thematic analysis resulted in five main themes, each with sub-themes. The full list of questions, themes, sub-themes, and participant votes are shown in Table \ref{table:questions1} in the Appendix. We note that some questions may belong to more than one (sub-)theme, however for our analysis, we chose the most representative classification. Themes and sub-themes, with examples selected mainly based on participant votes, are presented below: 

\subsection{Theme 1: Ethical, Legal, and Policy Considerations}

A primary theme (N=41) that emerged concerns the ethical, legal, and policy aspects of AI disclosures. This had three inter-related sub-themes. The first sub-theme covers the \textbf{Ethical Implications of AI Use}: \quoting[Q1]{How much should you communicate? Does law provide any answers? If not, what is ethical?}; \quoting[Q2]{How dangerous is the content generated by AI?}; \quoting[Q3]{Why should this be a right?}. The second sub-theme consisted of questions pertaining to \textbf{Legal Compliance and AI Disclosure}: \quoting[Q7]{What should be disclosed?}; \quoting[Q8]{Who is going to be responsible for the consequences?}; \quoting[Q12]{When should users be punished for not following the obligation? and when providers?}. The last sub-theme here consisted of questions related to \textbf{Policy and Regulatory Impact}. While some overlap may exist with legal compliance, questions under this sub-theme focused largely on the impact of such policies, rather than compliance per se. Examples here include: \quoting[Q23]{Who is going to be impacted by non disclosure?}; \quoting[Q24]{When do users need to disclose the use of AI? Consider continued influence effect}; \quoting[Q25]{Who decides if a ``context of use" is obvious enough to not require an ``interaction"?}. These sub-themes show how disclosures (and their effectiveness) are dependent on a variety of inter-related factors, the resolution of which, too has ethical and legal implications (e.g., if disclosed, what should be disclosed and who should do so; responsibility being also legally defined, etc.). This has prompted recent work to discuss what GenAI regulation should focus on (cf., \cite{Helberger2023-ac,Hacker2023,Novelli2024generative}) and how such measures can be enforced -- for example, given their multi-purpose usage, such general-purpose AI systems should consider safety from the onset, starting with data quality \cite{Helberger2023-ac}.

\subsection{Theme 2: Future Considerations, Evolving Context, \& Practical Implementation}

Another theme that emerged (N=19) concerns the future of AI technology and disclosures, their implementation, and their impact on society. This also had three sub-themes. The first sub-theme consisted of questions pertaining to the \textbf{Evolving AI Technologies and Societal Impact}, where key questions touched upon societal aspects where misinformation may be rampant, and included: \quoting[Q42]{Who cares?}; \quoting[Q45]{When can AI generated content affect real life affairs?}; \quoting[Q48]{Where could fake AI content show up?}. The second sub-theme consisted of questions covering \textbf{Future Trends and Legal Adaptation}: \quoting[Q49]{Why is it important to inform people?}; \quoting[Q50]{Why should people care about this?}; \quoting[Q52]{What is the definition of AI generated content? (e.g. images, text)}. The final sub-theme addresses questions related to \textbf{Practical Challenges in AI Implementation}: \quoting[Q56]{How much effort should be put into the classification?}; \quoting[Q58]{What kind of AI model could be used in such situations?}; \quoting[Q61]{How to trigger user feedback for AI disclosure or interactions?}. Such considerations are important for better understanding the impact on society. For example, Yaqub et al. \cite{Yaqub2020} confirmed that credibility indicators on social media can decrease the propensity to share fake news, however their impact varied, with fact checking services found to be most effective. Yet social media platforms come and go, with an evolving user base, which underscores the importance of accounting for dynamic (future) contexts.

\subsection{Theme 3: Provider Responsibility and Industry Impact}

Given that we focused on differences between users and providers, a theme (N=18) that emerged concerned the ethical responsibilities and obligations of the (media) provider. This consisted of three sub-themes. The first sub-theme concerns \textbf{Ethical Considerations for Providers}: \quoting[Q62]{Why do providers need to disclose they use AI?}; \quoting[Q64]{How should we treat users when not following the obligation? And how providers? (think punishment)}; \quoting[Q68]{Why providers should consider disclosure in their decision making process?}. The second sub-theme covers \textbf{Industry-Specific Impact and Challenges}: \quoting[Q68]{Why do providers want to use AI? (what is the benefit?)}; \quoting[Q71]{What industry will be affected the most?}; \quoting[Q72]{What is considered manipulation? Is there also a degree to this that is accounted for?}. The last sub-theme addresses \textbf{Provider Disclosure Obligations}: \quoting[Q74]{Where to label information / content?}; \quoting[Q77]{What freedom shall the provider have in formulating/offering information to the recipient?}; \quoting[Q78]{What information should the provider disclose to the recipient of AI content?}. Given these, we highlight not just the ethical but also the business impact of AI disclosures, which may itself then impact ethical considerations. For example, in a field customer purchase setting using chatbots, Luo et al. \cite{Luo2019} found that disclosure of chatbot identity before the machine--customer conversation took place reduced purchase rates by more than 79\%, despite what is known about people's machine heuristic \cite{Sundar2019}.

\subsection{Theme 4: Trust, Authenticity, and User Empowerment}

A key theme (N=33) of our workshops was on trust, authenticity, and how users can be empowered. This consisted of four sub-themes. The first sub-theme includes questions on \textbf{Authenticity, Provenance, and Transparency}: \quoting[Q81]{How important is it for users to know the authenticity of media?}; \quoting[Q82]{How are we going to verify authenticity?}; \quoting[Q84]{Where can people manually verify if it is AI generated?}. The second sub-theme tackles the Human-centered AI issue of \textbf{Building Trust in AI Systems}: \quoting[Q89]{How will different users interpret signals differently?}; \quoting[Q91]{How does disclosure impact people's trust and perceived credibility?}; \quoting[Q92]{How to responsibly disclose artistic uses of AI?}. The third sub-theme focuses on \textbf{Empowering Users through Education \& Awareness}: \quoting[Q99]{How can we train users to spot authentic content without relying on explicit cues?}; \quoting[Q100]{What type of information empowers/triggers ``act of resistance"?}; \quoting[Q103]{Where should we get ask if we suspect things (images, texts) are generated by AI?}. Finally, the last sub-theme focuses on \textbf{User Responsibility and Agency}: \quoting[Q110]{Why do users need to know they are interacting with AI?}; \quoting[Q111]{What information should the users acquire? Are they responsible for it?}; \quoting[Q112]{Why is it important for promoting democracy?}. The sub-themes here touch upon important prior work that aims to address authenticity and trust, in the context of media. Liao \& Sundar \cite{liao_designing_2022} introduced a conceptual model called MATCH that describes how trustworthiness is communicated in AI systems through trustworthiness cues, and propose a requirement checklist to support technology creators in identifying reliable cues. Scharowski et al. \cite{scharowski_certification_2023} examined the potential for AI certification labels (e.g., the "Digital Trust Label" by the 2023 Swiss Digital Initiative), and found that these can mitigate data-related concerns expressed by end-users (e.g., privacy and data protection), however other concerns (e.g., model performance) remain challenging to address. Longoni et al. \cite{Longoni2022} found that people rated news headlines written by AI as less accurate than those written by humans. Recently, Toff \& Simon \cite{Toff2023-qq} found that on average audiences perceive news labeled as AI-generated as less trustworthy, even when the articles themselves are not evaluated as any less accurate or unfair. However, these effects largely stem from those whose pre-existing levels of trust in news are higher to begin with and among those knowledgeable about journalism.

\subsection{Theme 5: User Experience, Information Overload, and Personalization}

Our final theme (N=34) concerned all matters related to User Experience (UX), information overload, and personalization. The first sub-theme concerns questions on \textbf{Personalization and User Preferences}: \quoting[Q115]{How much freedom should the users get to (not) see the classification?}; \quoting[Q117]{How to adapt AI disclosures personally?}; \quoting[Q118]{How to present them in different devices?}. The second sub-theme addresses questions on the \textbf{Psychology of Human-AI Interaction}: \quoting[Q120]{Why do users trust AI more/less than a human?}; \quoting[Q121]{How would the disclosure affect the users?}; \quoting[Q122]{What is the psychological difference between AI and Human?}. The third sub-theme covers questions on \textbf{Standardization}: \quoting[Q126]{What is reliable method to test the classification?}; \quoting[Q127]{Where on the website should Al be disclosed? Terms of service, every piece of content, every element?}; \quoting[Q128]{Who determines how disclosure should be provided, according to which standards?}. The fourth sub-theme focuses on \textbf{User Interfaces and Information Overload}: \quoting[Q133]{Where is the classification going to be displayed?}; \quoting[Q134]{Where is the balance between communicating enough and information overload?}; \quoting[Q135]{How can you effectively communicate use of Al without distracting from content?}. The final sub-theme concerns \textbf{User-Centric Information Design}: \quoting[Q139]{Who are you communicating to? e.g., different people may require different types of communication mechanisms}; \quoting[Q140]{What (under-) information is meaningful for the recipient?}; \quoting[Q142]{Where to provide the AI disclosure? Before the generated content or after that?}. Indeed, with respect to UX and personalization aspects of disclosure, the level of granularity by which such disclosures are shown (e.g., do they pertain to the training data or to surface interaction), any corresponding explanations, and the extent to which these increase trust, can further compound public risk perceptions. In this direction, in studying progressive disclosures for algorithmic decisions made by intelligent systems, Springer \& Whittaker \cite{springer_progressive_2020} found that users may benefit from initially simplified feedback that hides potential system errors, and thereafter assists users in building working system operation heuristics. This is line with Muraldihar et al.'s \cite{Muralidhar2023} survey that found a simplified explanation of the AI system is sufficient for ensuring transparency, allow users to have a better sense of system accuracy, fairness, and privacy.

\section{Discussion and Conclusion}

\subsection{Limitations and future work}

First, AI technology is advancing rapidly, and our participant responses were likely geared toward their experiences with GenAI technology today. However in the future, more transparent and robust AI systems may afford different questions than what we observed. Relatedly, the law is not immutable -- this is continuously evolving, and the AI Act may develop to cover some of the key questions raised in this work. Lastly, while we took care to ensure our question set and themes are exhaustive for the study of AI disclosures, there may be aspects that were missed. Similarly, themes may be decomposed further -- however, for our current scope, we believe these themes have sufficient coverage and detail to inform future work. Lastly, these should not be seen as a final but rather organic, ever-growing list of questions to systematically guide research on transparent AI disclosures.

\subsection{Transparent AI disclosures in and beyond the AI Act}

Our work aimed at tackling the challenge of interpreting and implementing transparent AI disclosures for continuously evolving AI technology, through the lens of Article 52 as currently stated in the draft EU AI Act \cite{EUAIAct}. Through the 5W1H questions, we arrived at several themes and sub-themes -- we believe each poses unique challenges for users, providers, and the media sector at large. With GenAI, media organizations are or will be undergoing a fundamental shift in practices. To echo Rakova et al. \cite{Rakova2021responsibleai}, to better enable responsible AI work, organizations need to update their practices, which will require addressing both prevalent and emerging work practices. For transparent AI disclosures, as we observed from the diverse set of questions tackling multiple aspects of disclosures (from terminology to ethical and legal aspects to societal impact), we believe this is a complex, multi-faceted challenge (a `wicked' problem) that requires concerted interdisciplinary efforts from researchers and practitioners across industries and institutes. As such, we believe on continually drawing on participatory AI and value-sensitive design approaches \cite{Birhane2022,Ehsan2020,Friedman1996} to creating AI for social good \cite{Umbrello2021valuedesign}, ensuring human creative practices are safeguarded and can flourish (cf., \cite{li2024user,inie2023designing}). This means ensuring media organization practices and citizens (end users) are transparently informed when they interact with AI-generated media in a meaningful, usable manner, throughout the human-AI interaction timeline. As a step toward this change, we believe our contributed key questions drawn from wide-varying expertise can serve as a practical starting point for cross-discipline research in AI disclosure -- by knowing what questions to ask, and for whom such questions may be most relevant, collaborative efforts can be strengthened.

Whether the most effective mechanism to tackle transparent AI disclosures constitutes watermarking to ensure copyright and traceable accountability \cite{tohline2008computational,Zhong2023Copyright}, or the continued refinement of regulation and legal policy at an (inter-)national level (of which the AI Act is a strong pillar \cite{VealeAIAct2021}), it is clear GenAI development needs to be acted upon responsibly - for and beyond media organization practice. For example, consider the use and misuse behind data privacy and security measures through interaction with consent banners \cite{gray_dark_2021}, despite EU efforts such as in General Data Protection Regulation (GDPR) mandates. These can lend themselves to dark patterns and potentially malicious use of AI persuasion \cite{burtell_artificial_2023}, raising serious ethical concerns pertaining to perceptual manipulations, cybersecurity, and the global risk of misinformation (cf., WEF's 2024 Global Risks Report \cite{WEF2024}). To this end, we believe drawing on key human-centered principles (e.g., human well-being alignment, responsible design of transparent AI) of responsible AI development \cite{Garibay2023SixChallengesHCAI} are of immediate necessity to protect citizens moving forward, and ensuring a well functioning democratic society.

\subsection{Conclusion}

Our work contributes 149 questions clustered into five themes and 18 sub-themes, that we believe can assist in tackling the challenge of transparently communicating AI disclosures -- for media providers, as well as end-users consuming media content everyday across platforms and devices. We hope our questions underscore the importance of better understanding user needs and reactions to transparency obligations, and support establishing user-centric designs for transparent AI disclosures that ultimately foster democratic societies based on truth rather than AI-generated fiction.

\begin{acks}

We thank all our participants for participating in our workshops, and acknowledge their valuable contributions to this early work. This publication is part of the AI, Media \& Democracy Lab (Dutch Research Council project number: NWA.1332.20.009).

\end{acks}

\bibliographystyle{ACM-Reference-Format}
\interlinepenalty=10000
\bibliography{ai_disclosure}


\begin{thebibliography}{59}


\ifx \showCODEN    \undefined \def \showCODEN     #1{\unskip}     \fi
\ifx \showDOI      \undefined \def \showDOI       #1{#1}\fi
\ifx \showISBNx    \undefined \def \showISBNx     #1{\unskip}     \fi
\ifx \showISBNxiii \undefined \def \showISBNxiii  #1{\unskip}     \fi
\ifx \showISSN     \undefined \def \showISSN      #1{\unskip}     \fi
\ifx \showLCCN     \undefined \def \showLCCN      #1{\unskip}     \fi
\ifx \shownote     \undefined \def \shownote      #1{#1}          \fi
\ifx \showarticletitle \undefined \def \showarticletitle #1{#1}   \fi
\ifx \showURL      \undefined \def \showURL       {\relax}        \fi
\providecommand\bibfield[2]{#2}
\providecommand\bibinfo[2]{#2}
\providecommand\natexlab[1]{#1}
\providecommand\showeprint[2][]{arXiv:#2}

\bibitem[\protect\citeauthoryear{2024}{2024}{2024}]%
        {WEF2024}
\bibfield{author}{\bibinfo{person}{World Economic~Forum 2024}.}
  \bibinfo{year}{2024}\natexlab{}.
\newblock \bibinfo{title}{The Global Risks Report 2024}.
\newblock
  \bibinfo{howpublished}{\url{https://www.weforum.org/publications/global-risks-report-2024}}.
\newblock
\newblock
\shownote{[2024-1-19].}


\bibitem[\protect\citeauthoryear{ACM}{ACM}{2024}]%
        {ACMauthorship}
\bibfield{author}{\bibinfo{person}{ACM}.} \bibinfo{year}{2024}\natexlab{}.
\newblock \bibinfo{title}{{A}{C}{M} {P}olicy on {A}uthorship}.
\newblock
  \bibinfo{howpublished}{\url{https://www.acm.org/publications/policies/new-acm-policy-on-authorship}}.
\newblock
\newblock
\shownote{[Accessed 22-01-2024].}


\bibitem[\protect\citeauthoryear{Act}{Act}{2022}]%
        {EUAIAct}
\bibfield{author}{\bibinfo{person}{European~AI Act}.}
  \bibinfo{year}{2022}\natexlab{}.
\newblock \bibinfo{title}{Proposal for a Regulation of the European Parliament
  and of the Council laying down harmonised rules on artificial intelligence
  (Artificial Intelligence Act) and amending certain Union legislative acts)
  and amending certain Union legislative acts}.
\newblock
  \bibinfo{howpublished}{\url{https://eur-lex.europa.eu/legal-content/EN/TXT/?uri=CELEX\%3A52021PC0206}}.
\newblock
\urldef\tempurl%
\url{https://data.consilium.europa.eu/doc/document/ST-14954-2022-INIT/en/pdf}
\showURL{%
\tempurl}
\newblock
\shownote{[Accessed 12-01-2024].}


\bibitem[\protect\citeauthoryear{Birhane, Isaac, Prabhakaran, Diaz, Elish,
  Gabriel, and Mohamed}{Birhane et~al\mbox{.}}{2022}]%
        {Birhane2022}
\bibfield{author}{\bibinfo{person}{Abeba Birhane}, \bibinfo{person}{William
  Isaac}, \bibinfo{person}{Vinodkumar Prabhakaran}, \bibinfo{person}{Mark
  Diaz}, \bibinfo{person}{Madeleine~Clare Elish}, \bibinfo{person}{Iason
  Gabriel}, {and} \bibinfo{person}{Shakir Mohamed}.}
  \bibinfo{year}{2022}\natexlab{}.
\newblock \showarticletitle{Power to the People? Opportunities and Challenges
  for Participatory AI}. In \bibinfo{booktitle}{\emph{Equity and Access in
  Algorithms, Mechanisms, and Optimization}} (Arlington, VA, USA)
  \emph{(\bibinfo{series}{EAAMO '22})}. \bibinfo{publisher}{Association for
  Computing Machinery}, \bibinfo{address}{New York, NY, USA}, Article
  \bibinfo{articleno}{6}, \bibinfo{numpages}{8}~pages.
\newblock
\showISBNx{9781450394772}
\urldef\tempurl%
\url{https://doi.org/10.1145/3551624.3555290}
\showDOI{\tempurl}


\bibitem[\protect\citeauthoryear{Bommasani, Hudson, Adeli, Altman, Arora, von
  Arx, Bernstein, Bohg, Bosselut, et~al\mbox{.}}{Bommasani
  et~al\mbox{.}}{2021}]%
        {Bommasani2021FoundationModels}
\bibfield{author}{\bibinfo{person}{Rishi Bommasani}, \bibinfo{person}{Drew~A.
  Hudson}, \bibinfo{person}{Ehsan Adeli}, \bibinfo{person}{Russ Altman},
  \bibinfo{person}{Simran Arora}, \bibinfo{person}{Sydney von Arx},
  \bibinfo{person}{Michael~S. Bernstein}, \bibinfo{person}{Jeannette Bohg},
  \bibinfo{person}{Antoine Bosselut}, {et~al\mbox{.}}}
  \bibinfo{year}{2021}\natexlab{}.
\newblock \showarticletitle{On the Opportunities and Risks of Foundation
  Models}.
\newblock \bibinfo{journal}{\emph{arXiv preprint arXiv:2108.07258}}
  (\bibinfo{year}{2021}).
\newblock
\urldef\tempurl%
\url{https://crfm.stanford.edu/assets/report.pdf}
\showURL{%
\tempurl}


\bibitem[\protect\citeauthoryear{Borsos, Marinier, Vincent, Kharitonov,
  Pietquin, Sharifi, Teboul, Grangier, Tagliasacchi, and Zeghidour}{Borsos
  et~al\mbox{.}}{2022}]%
        {borsos2022audiolm}
\bibfield{author}{\bibinfo{person}{Zal\'{a}n Borsos},
  \bibinfo{person}{Rapha\"{e}l Marinier}, \bibinfo{person}{Damien Vincent},
  \bibinfo{person}{Eugene Kharitonov}, \bibinfo{person}{Olivier Pietquin},
  \bibinfo{person}{Matt Sharifi}, \bibinfo{person}{Olivier Teboul},
  \bibinfo{person}{David Grangier}, \bibinfo{person}{Marco Tagliasacchi}, {and}
  \bibinfo{person}{Neil Zeghidour}.} \bibinfo{year}{2022}\natexlab{}.
\newblock \bibinfo{title}{AudioLM: a Language Modeling Approach to Audio
  Generation}.
\newblock
\newblock
\showeprint[arxiv]{2209.03143}~[cs.SD]


\bibitem[\protect\citeauthoryear{Burtell and Woodside}{Burtell and
  Woodside}{2023}]%
        {burtell_artificial_2023}
\bibfield{author}{\bibinfo{person}{Matthew Burtell} {and}
  \bibinfo{person}{Thomas Woodside}.} \bibinfo{year}{2023}\natexlab{}.
\newblock \bibinfo{title}{Artificial {Influence}: {An} {Analysis} {Of}
  {AI}-{Driven} {Persuasion}}.
\newblock
\newblock
\urldef\tempurl%
\url{http://arxiv.org/abs/2303.08721}
\showURL{%
\tempurl}
\newblock
\shownote{arXiv:2303.08721 [cs].}


\bibitem[\protect\citeauthoryear{C2PA}{C2PA}{2024}]%
        {c2paIntroducingOfficial}
\bibfield{author}{\bibinfo{person}{C2PA}.} \bibinfo{year}{2024}\natexlab{}.
\newblock \bibinfo{title}{{I}ntroducing {O}fficial {C}ontent {C}redentials
  {I}con - {C}2{P}{A} --- c2pa.org}.
\newblock
  \bibinfo{howpublished}{\url{https://c2pa.org/post/contentcredentials/}}.
\newblock
\newblock
\shownote{[Accessed 17-01-2024].}


\bibitem[\protect\citeauthoryear{Clarke, Braun, and Hayfield}{Clarke
  et~al\mbox{.}}{2015}]%
        {clarke2015thematic}
\bibfield{author}{\bibinfo{person}{Victoria Clarke}, \bibinfo{person}{Virginia
  Braun}, {and} \bibinfo{person}{Nikki Hayfield}.}
  \bibinfo{year}{2015}\natexlab{}.
\newblock \showarticletitle{Thematic analysis}.
\newblock \bibinfo{journal}{\emph{Qualitative psychology: A practical guide to
  research methods}} \bibinfo{volume}{222}, \bibinfo{number}{2015}
  (\bibinfo{year}{2015}), \bibinfo{pages}{248}.
\newblock


\bibitem[\protect\citeauthoryear{Cloudy, Banks, and Bowman}{Cloudy
  et~al\mbox{.}}{2023}]%
        {cloudy_straight_2023}
\bibfield{author}{\bibinfo{person}{Joshua Cloudy}, \bibinfo{person}{Jaime
  Banks}, {and} \bibinfo{person}{Nicholas~David Bowman}.}
  \bibinfo{year}{2023}\natexlab{}.
\newblock \showarticletitle{The {Str}({AI})ght {Scoop}: {Artificial}
  {Intelligence} {Cues} {Reduce} {Perceptions} of {Hostile} {Media} {Bias}}.
\newblock \bibinfo{journal}{\emph{Digital Journalism}} \bibinfo{volume}{11},
  \bibinfo{number}{9} (\bibinfo{date}{Oct.} \bibinfo{year}{2023}),
  \bibinfo{pages}{1577--1596}.
\newblock
\showISSN{2167-0811, 2167-082X}
\urldef\tempurl%
\url{https://doi.org/10.1080/21670811.2021.1969974}
\showDOI{\tempurl}


\bibitem[\protect\citeauthoryear{{DeepMind}}{{DeepMind}}{2024}]%
        {SynthID}
\bibfield{author}{\bibinfo{person}{Google {DeepMind}}.}
  \bibinfo{year}{2024}\natexlab{}.
\newblock \bibinfo{title}{{SynthID}}.
\newblock
  \bibinfo{howpublished}{\url{https://deepmind.google/technologies/synthid/}}.
\newblock
\newblock
\shownote{Accessed: 2024-1-19.}


\bibitem[\protect\citeauthoryear{Ehsan and Riedl}{Ehsan and Riedl}{2020}]%
        {Ehsan2020}
\bibfield{author}{\bibinfo{person}{Upol Ehsan} {and} \bibinfo{person}{Mark~O.
  Riedl}.} \bibinfo{year}{2020}\natexlab{}.
\newblock \showarticletitle{Human-Centered Explainable AI: Towards a Reflective
  Sociotechnical Approach}. In \bibinfo{booktitle}{\emph{HCI International 2020
  - Late Breaking Papers: Multimodality and Intelligence}},
  \bibfield{editor}{\bibinfo{person}{Constantine Stephanidis},
  \bibinfo{person}{Masaaki Kurosu}, \bibinfo{person}{Helmut Degen}, {and}
  \bibinfo{person}{Lauren Reinerman-Jones}} (Eds.).
  \bibinfo{publisher}{Springer International Publishing},
  \bibinfo{address}{Cham}, \bibinfo{pages}{449--466}.
\newblock
\showISBNx{978-3-030-60117-1}


\bibitem[\protect\citeauthoryear{Elagroudy, Li, V\"{a}\"{a}n\"{a}nen, Lukowicz,
  Ishii, Mackay, Churchill, Peters, Oulasvirta, Prada, Diening, Barbareschi,
  Gruenerbl, Kawaguchi, Ali, Draxler, Welsch, and Schmidt}{Elagroudy
  et~al\mbox{.}}{2024}]%
        {Elagroudy2024}
\bibfield{author}{\bibinfo{person}{Passant Elagroudy}, \bibinfo{person}{Jie
  Li}, \bibinfo{person}{Kaisa V\"{a}\"{a}n\"{a}nen}, \bibinfo{person}{Paul
  Lukowicz}, \bibinfo{person}{Hiroshi Ishii}, \bibinfo{person}{Wendy Mackay},
  \bibinfo{person}{Elizabeth Churchill}, \bibinfo{person}{Anicia Peters},
  \bibinfo{person}{Antti Oulasvirta}, \bibinfo{person}{Rui Prada},
  \bibinfo{person}{Alexandra Diening}, \bibinfo{person}{Giulia Barbareschi},
  \bibinfo{person}{Agnes Gruenerbl}, \bibinfo{person}{Midori Kawaguchi},
  \bibinfo{person}{Abdallah~El Ali}, \bibinfo{person}{Fiona Draxler},
  \bibinfo{person}{Robin Welsch}, {and} \bibinfo{person}{Albrecht Schmidt}.}
  \bibinfo{year}{2024}\natexlab{}.
\newblock \showarticletitle{Transforming HCI Research Cycles using Generative
  AI and ``Large Whatever Models" (LWMs)}. In
  \bibinfo{booktitle}{\emph{Extended Abstracts of the 2024 CHI Conference on
  Human Factors in Computing Systems}} (Honolulu, HI, USA)
  \emph{(\bibinfo{series}{CHI '24 EA})}. \bibinfo{publisher}{Association for
  Computing Machinery}, \bibinfo{address}{New York, NY, USA},
  \bibinfo{numpages}{5}~pages.
\newblock
\showISBNx{979-8-4007-0331-7/24/05}
\urldef\tempurl%
\url{https://doi.org/10.1145/3613905.3643977}
\showDOI{\tempurl}


\bibitem[\protect\citeauthoryear{Epstein, Fang, Arechar, and Rand}{Epstein
  et~al\mbox{.}}{2023}]%
        {Epstein2023}
\bibfield{author}{\bibinfo{person}{Ziv Epstein}, \bibinfo{person}{Mengying~C
  Fang}, \bibinfo{person}{Antonio~A Arechar}, {and} \bibinfo{person}{David~G
  Rand}.} \bibinfo{year}{2023}\natexlab{}.
\newblock \bibinfo{title}{What label should be applied to content produced by
  generative AI?}
\newblock
\newblock
\urldef\tempurl%
\url{https://doi.org/10.31234/osf.io/v4mfz}
\showDOI{\tempurl}


\bibitem[\protect\citeauthoryear{Friedman}{Friedman}{1996}]%
        {Friedman1996}
\bibfield{author}{\bibinfo{person}{Batya Friedman}.}
  \bibinfo{year}{1996}\natexlab{}.
\newblock \showarticletitle{Value-Sensitive Design}.
\newblock \bibinfo{journal}{\emph{Interactions}} \bibinfo{volume}{3},
  \bibinfo{number}{6} (\bibinfo{date}{dec} \bibinfo{year}{1996}),
  \bibinfo{pages}{16--23}.
\newblock
\showISSN{1072-5520}
\urldef\tempurl%
\url{https://doi.org/10.1145/242485.242493}
\showDOI{\tempurl}


\bibitem[\protect\citeauthoryear{Garibay, Winslow, Andolina, Antona,
  Bodenschatz, Coursaris, Falco, Fiore, Garibay, Grieman, Havens, Jirotka,
  Kacorri, Karwowski, Kider, Konstan, Koon, Lopez-Gonzalez, Maifeld-Carucci,
  McGregor, Salvendy, Shneiderman, Stephanidis, Strobel, Holter, and
  Xu}{Garibay et~al\mbox{.}}{2023}]%
        {Garibay2023SixChallengesHCAI}
\bibfield{author}{\bibinfo{person}{Ozlem~Ozmen Garibay}, \bibinfo{person}{Brent
  Winslow}, \bibinfo{person}{Salvatore Andolina}, \bibinfo{person}{Margherita
  Antona}, \bibinfo{person}{Anja Bodenschatz}, \bibinfo{person}{Constantinos
  Coursaris}, \bibinfo{person}{Gregory Falco}, \bibinfo{person}{Stephen~M.
  Fiore}, \bibinfo{person}{Ivan Garibay}, \bibinfo{person}{Keri Grieman},
  \bibinfo{person}{John~C. Havens}, \bibinfo{person}{Marina Jirotka},
  \bibinfo{person}{Hernisa Kacorri}, \bibinfo{person}{Waldemar Karwowski},
  \bibinfo{person}{Joe Kider}, \bibinfo{person}{Joseph Konstan},
  \bibinfo{person}{Sean Koon}, \bibinfo{person}{Monica Lopez-Gonzalez},
  \bibinfo{person}{Iliana Maifeld-Carucci}, \bibinfo{person}{Sean McGregor},
  \bibinfo{person}{Gavriel Salvendy}, \bibinfo{person}{Ben Shneiderman},
  \bibinfo{person}{Constantine Stephanidis}, \bibinfo{person}{Christina
  Strobel}, \bibinfo{person}{Carolyn~Ten Holter}, {and} \bibinfo{person}{Wei
  Xu}.} \bibinfo{year}{2023}\natexlab{}.
\newblock \showarticletitle{Six Human-Centered Artificial Intelligence Grand
  Challenges}.
\newblock \bibinfo{journal}{\emph{International Journal of Human Computer
  Interaction}} \bibinfo{volume}{39}, \bibinfo{number}{3}
  (\bibinfo{year}{2023}), \bibinfo{pages}{391--437}.
\newblock
\urldef\tempurl%
\url{https://doi.org/10.1080/10447318.2022.2153320}
\showDOI{\tempurl}
\showeprint{https://doi.org/10.1080/10447318.2022.2153320}


\bibitem[\protect\citeauthoryear{Gray, Santos, Bielova, Toth, and
  Clifford}{Gray et~al\mbox{.}}{2021}]%
        {gray_dark_2021}
\bibfield{author}{\bibinfo{person}{Colin~M. Gray}, \bibinfo{person}{Cristiana
  Santos}, \bibinfo{person}{Nataliia Bielova}, \bibinfo{person}{Michael Toth},
  {and} \bibinfo{person}{Damian Clifford}.} \bibinfo{year}{2021}\natexlab{}.
\newblock \showarticletitle{Dark {Patterns} and the {Legal} {Requirements} of
  {Consent} {Banners}: {An} {Interaction} {Criticism} {Perspective}}. In
  \bibinfo{booktitle}{\emph{Proc. CHI '21}}. \bibinfo{publisher}{ACM},
  \bibinfo{address}{Yokohama Japan}, \bibinfo{pages}{1--18}.
\newblock
\showISBNx{978-1-4503-8096-6}
\urldef\tempurl%
\url{https://doi.org/10.1145/3411764.3445779}
\showDOI{\tempurl}


\bibitem[\protect\citeauthoryear{Groh, Sankaranarayanan, Singh, Kim, Lippman,
  and Picard}{Groh et~al\mbox{.}}{2023}]%
        {groh2023human}
\bibfield{author}{\bibinfo{person}{Matthew Groh}, \bibinfo{person}{Aruna
  Sankaranarayanan}, \bibinfo{person}{Nikhil Singh},
  \bibinfo{person}{Dong~Young Kim}, \bibinfo{person}{Andrew Lippman}, {and}
  \bibinfo{person}{Rosalind Picard}.} \bibinfo{year}{2023}\natexlab{}.
\newblock \bibinfo{title}{Human Detection of Political Speech Deepfakes across
  Transcripts, Audio, and Video}.
\newblock
\newblock
\showeprint[arxiv]{2202.12883}~[cs.HC]


\bibitem[\protect\citeauthoryear{Hacker, Engel, and Mauer}{Hacker
  et~al\mbox{.}}{2023}]%
        {Hacker2023}
\bibfield{author}{\bibinfo{person}{Philipp Hacker}, \bibinfo{person}{Andreas
  Engel}, {and} \bibinfo{person}{Marco Mauer}.}
  \bibinfo{year}{2023}\natexlab{}.
\newblock \showarticletitle{Regulating ChatGPT and Other Large Generative AI
  Models}. In \bibinfo{booktitle}{\emph{Proceedings of the 2023 ACM Conference
  on Fairness, Accountability, and Transparency}} (Chicago, IL, USA)
  \emph{(\bibinfo{series}{FAccT '23})}. \bibinfo{publisher}{Association for
  Computing Machinery}, \bibinfo{address}{New York, NY, USA},
  \bibinfo{pages}{1112?1123}.
\newblock
\showISBNx{9798400701924}
\urldef\tempurl%
\url{https://doi.org/10.1145/3593013.3594067}
\showDOI{\tempurl}


\bibitem[\protect\citeauthoryear{Hart}{Hart}{1996}]%
        {Hart1996}
\bibfield{author}{\bibinfo{person}{Geoff Hart}.}
  \bibinfo{year}{1996}\natexlab{}.
\newblock \showarticletitle{The Five W's: An Old Tool for the New Task of
  Audience Analysis}.
\newblock \bibinfo{journal}{\emph{Technical Communication}}
  \bibinfo{volume}{43}, \bibinfo{number}{2} (\bibinfo{year}{1996}),
  \bibinfo{pages}{139--145}.
\newblock
\showISSN{00493155, 1938369X}
\urldef\tempurl%
\url{http://www.jstor.org/stable/43088033}
\showURL{%
\tempurl}


\bibitem[\protect\citeauthoryear{Helberger and Diakopoulos}{Helberger and
  Diakopoulos}{2023}]%
        {Helberger2023-ac}
\bibfield{author}{\bibinfo{person}{Natali Helberger} {and}
  \bibinfo{person}{Nicholas Diakopoulos}.} \bibinfo{year}{2023}\natexlab{}.
\newblock \showarticletitle{{ChatGPT} and the {AI} act}.
\newblock \bibinfo{journal}{\emph{Internet Pol. Rev.}} \bibinfo{volume}{12},
  \bibinfo{number}{1} (\bibinfo{date}{Feb.} \bibinfo{year}{2023}).
\newblock


\bibitem[\protect\citeauthoryear{Ho, Chan, Saharia, Whang, Gao, Gritsenko,
  Kingma, Poole, Norouzi, Fleet, et~al\mbox{.}}{Ho et~al\mbox{.}}{2022}]%
        {ho2022imagenvideo}
\bibfield{author}{\bibinfo{person}{Jonathan Ho}, \bibinfo{person}{William
  Chan}, \bibinfo{person}{Chitwan Saharia}, \bibinfo{person}{Jay Whang},
  \bibinfo{person}{Ruiqi Gao}, \bibinfo{person}{Alexey Gritsenko},
  \bibinfo{person}{Diederik~P Kingma}, \bibinfo{person}{Ben Poole},
  \bibinfo{person}{Mohammad Norouzi}, \bibinfo{person}{David~J Fleet},
  {et~al\mbox{.}}} \bibinfo{year}{2022}\natexlab{}.
\newblock \bibinfo{title}{Imagen video: High definition video generation with
  diffusion models}.
\newblock
\newblock
\showeprint[arxiv]{2210.02303}~[cs.CV]


\bibitem[\protect\citeauthoryear{Hosseini, Resnik, and Holmes}{Hosseini
  et~al\mbox{.}}{2023}]%
        {Hosseini2023}
\bibfield{author}{\bibinfo{person}{Mohammad Hosseini}, \bibinfo{person}{David~B
  Resnik}, {and} \bibinfo{person}{Kristi Holmes}.}
  \bibinfo{year}{2023}\natexlab{}.
\newblock \showarticletitle{The ethics of disclosing the use of artificial
  intelligence tools in writing scholarly manuscripts}.
\newblock \bibinfo{journal}{\emph{Research Ethics}} \bibinfo{volume}{19},
  \bibinfo{number}{4} (\bibinfo{year}{2023}), \bibinfo{pages}{449--465}.
\newblock
\urldef\tempurl%
\url{https://doi.org/10.1177/17470161231180449}
\showDOI{\tempurl}
\showeprint{https://doi.org/10.1177/17470161231180449}


\bibitem[\protect\citeauthoryear{Inie, Falk, and Tanimoto}{Inie
  et~al\mbox{.}}{2023}]%
        {inie2023designing}
\bibfield{author}{\bibinfo{person}{Nanna Inie}, \bibinfo{person}{Jeanette
  Falk}, {and} \bibinfo{person}{Steve Tanimoto}.}
  \bibinfo{year}{2023}\natexlab{}.
\newblock \showarticletitle{Designing Participatory AI: Creative
  Professionals’ Worries and Expectations about Generative AI}. In
  \bibinfo{booktitle}{\emph{Extended Abstracts of the 2023 CHI Conference on
  Human Factors in Computing Systems}}. \bibinfo{pages}{1--8}.
\newblock


\bibitem[\protect\citeauthoryear{Jia, Boltz, Zhang, Chen, and Lee}{Jia
  et~al\mbox{.}}{2022}]%
        {Jia2022}
\bibfield{author}{\bibinfo{person}{Chenyan Jia}, \bibinfo{person}{Alexander
  Boltz}, \bibinfo{person}{Angie Zhang}, \bibinfo{person}{Anqing Chen}, {and}
  \bibinfo{person}{Min~Kyung Lee}.} \bibinfo{year}{2022}\natexlab{}.
\newblock \showarticletitle{Understanding Effects of Algorithmic vs. Community
  Label on Perceived Accuracy of Hyper-Partisan Misinformation}.
\newblock \bibinfo{journal}{\emph{Proc. ACM Hum.-Comput. Interact.}}
  \bibinfo{volume}{6}, \bibinfo{number}{CSCW2}, Article
  \bibinfo{articleno}{371} (\bibinfo{date}{nov} \bibinfo{year}{2022}),
  \bibinfo{numpages}{27}~pages.
\newblock
\urldef\tempurl%
\url{https://doi.org/10.1145/3555096}
\showDOI{\tempurl}


\bibitem[\protect\citeauthoryear{Jia, Cai, Yu, and Tse}{Jia
  et~al\mbox{.}}{2016}]%
        {Jia2016}
\bibfield{author}{\bibinfo{person}{Changjiang Jia}, \bibinfo{person}{Yan Cai},
  \bibinfo{person}{Yuen~Tak Yu}, {and} \bibinfo{person}{T.H. Tse}.}
  \bibinfo{year}{2016}\natexlab{}.
\newblock \showarticletitle{5W+1H pattern: A perspective of systematic mapping
  studies and a case study on cloud software testing}.
\newblock \bibinfo{journal}{\emph{Journal of Systems and Software}}
  \bibinfo{volume}{116} (\bibinfo{year}{2016}), \bibinfo{pages}{206--219}.
\newblock
\showISSN{0164-1212}
\urldef\tempurl%
\url{https://doi.org/10.1016/j.jss.2015.01.058}
\showDOI{\tempurl}


\bibitem[\protect\citeauthoryear{Kernis and Goldman}{Kernis and
  Goldman}{2006}]%
        {kernis_multicomponent_2006}
\bibfield{author}{\bibinfo{person}{Michael~H. Kernis} {and}
  \bibinfo{person}{Brian~M. Goldman}.} \bibinfo{year}{2006}\natexlab{}.
\newblock \showarticletitle{A {Multicomponent} {Conceptualization} of
  {Authenticity}: {Theory} and {Research}}.
\newblock In \bibinfo{booktitle}{\emph{Advances in {Experimental} {Social}
  {Psychology}}}. Vol.~\bibinfo{volume}{38}. \bibinfo{publisher}{Elsevier},
  \bibinfo{pages}{283--357}.
\newblock
\showISBNx{978-0-12-015238-4}
\urldef\tempurl%
\url{https://doi.org/10.1016/S0065-2601(06)38006-9}
\showDOI{\tempurl}


\bibitem[\protect\citeauthoryear{Langer, Hunsicker, Feldkamp, K\"{o}nig, and
  Grgi\'{c}-Hla\v{c}a}{Langer et~al\mbox{.}}{2022}]%
        {Langer2022}
\bibfield{author}{\bibinfo{person}{Markus Langer}, \bibinfo{person}{Tim
  Hunsicker}, \bibinfo{person}{Tina Feldkamp}, \bibinfo{person}{Cornelius~J.
  K\"{o}nig}, {and} \bibinfo{person}{Nina Grgi\'{c}-Hla\v{c}a}.}
  \bibinfo{year}{2022}\natexlab{}.
\newblock \showarticletitle{``Look! It's a Computer Program! It's an Algorithm!
  It's AI!": Does Terminology Affect Human Perceptions and Evaluations of
  Algorithmic Decision-Making Systems?}. In
  \bibinfo{booktitle}{\emph{Proceedings of the 2022 CHI Conference on Human
  Factors in Computing Systems}} (New Orleans, LA, USA)
  \emph{(\bibinfo{series}{CHI '22})}. \bibinfo{publisher}{Association for
  Computing Machinery}, \bibinfo{address}{New York, NY, USA}, Article
  \bibinfo{articleno}{581}, \bibinfo{numpages}{28}~pages.
\newblock
\showISBNx{9781450391573}
\urldef\tempurl%
\url{https://doi.org/10.1145/3491102.3517527}
\showDOI{\tempurl}


\bibitem[\protect\citeauthoryear{Li, Cao, Lin, Hou, Zhu, and Ali}{Li
  et~al\mbox{.}}{2024}]%
        {li2024user}
\bibfield{author}{\bibinfo{person}{Jie Li}, \bibinfo{person}{Hancheng Cao},
  \bibinfo{person}{Laura Lin}, \bibinfo{person}{Youyang Hou},
  \bibinfo{person}{Ruihao Zhu}, {and} \bibinfo{person}{Abdallah~El Ali}.}
  \bibinfo{year}{2024}\natexlab{}.
\newblock \showarticletitle{User Experience Design Professionals' Perceptions
  of Generative Artificial Intelligence}. In
  \bibinfo{booktitle}{\emph{Proceedings of the 2024 CHI Conference on Human
  Factors in Computing Systems}} (Honolulu, HI, USA)
  \emph{(\bibinfo{series}{CHI '24})}. \bibinfo{publisher}{Association for
  Computing Machinery}, \bibinfo{address}{New York, NY, USA},
  \bibinfo{numpages}{15}~pages.
\newblock
\showISBNx{979-8-4007-0330-0/24/05}
\urldef\tempurl%
\url{https://doi.org/10.1145/3613904.3642114}
\showDOI{\tempurl}


\bibitem[\protect\citeauthoryear{Liang, Yuksekgonul, Mao, Wu, and Zou}{Liang
  et~al\mbox{.}}{2023}]%
        {Liang2023}
\bibfield{author}{\bibinfo{person}{Weixin Liang}, \bibinfo{person}{Mert
  Yuksekgonul}, \bibinfo{person}{Yining Mao}, \bibinfo{person}{Eric Wu}, {and}
  \bibinfo{person}{James Zou}.} \bibinfo{year}{2023}\natexlab{}.
\newblock \showarticletitle{GPT detectors are biased against non-native English
  writers}.
\newblock \bibinfo{journal}{\emph{Patterns}} \bibinfo{volume}{4},
  \bibinfo{number}{7} (\bibinfo{year}{2023}), \bibinfo{pages}{100779}.
\newblock
\showISSN{2666-3899}
\urldef\tempurl%
\url{https://doi.org/10.1016/j.patter.2023.100779}
\showDOI{\tempurl}


\bibitem[\protect\citeauthoryear{Liao and Sundar}{Liao and Sundar}{2022}]%
        {liao_designing_2022}
\bibfield{author}{\bibinfo{person}{Q.Vera Liao} {and} \bibinfo{person}{S.~Shyam
  Sundar}.} \bibinfo{year}{2022}\natexlab{}.
\newblock \showarticletitle{Designing for {Responsible} {Trust} in {AI}
  {Systems}: {A} {Communication} {Perspective}}. In
  \bibinfo{booktitle}{\emph{2022 {ACM} {Conference} on {Fairness},
  {Accountability}, and {Transparency}}}. \bibinfo{publisher}{ACM},
  \bibinfo{address}{Seoul Republic of Korea}, \bibinfo{pages}{1257--1268}.
\newblock
\showISBNx{978-1-4503-9352-2}
\urldef\tempurl%
\url{https://doi.org/10.1145/3531146.3533182}
\showDOI{\tempurl}


\bibitem[\protect\citeauthoryear{Longoni, Fradkin, Cian, and Pennycook}{Longoni
  et~al\mbox{.}}{2022}]%
        {Longoni2022}
\bibfield{author}{\bibinfo{person}{Chiara Longoni}, \bibinfo{person}{Andrey
  Fradkin}, \bibinfo{person}{Luca Cian}, {and} \bibinfo{person}{Gordon
  Pennycook}.} \bibinfo{year}{2022}\natexlab{}.
\newblock \showarticletitle{News from Generative Artificial Intelligence Is
  Believed Less}. In \bibinfo{booktitle}{\emph{Proceedings of the 2022 ACM
  Conference on Fairness, Accountability, and Transparency}} (Seoul, Republic
  of Korea) \emph{(\bibinfo{series}{FAccT '22})}.
  \bibinfo{publisher}{Association for Computing Machinery},
  \bibinfo{address}{New York, NY, USA}, \bibinfo{pages}{97--106}.
\newblock
\showISBNx{9781450393522}
\urldef\tempurl%
\url{https://doi.org/10.1145/3531146.3533077}
\showDOI{\tempurl}


\bibitem[\protect\citeauthoryear{Lukianets}{Lukianets}{2024}]%
        {openethicsTaxonomy}
\bibfield{author}{\bibinfo{person}{Nikita Lukianets}.}
  \bibinfo{year}{2024}\natexlab{}.
\newblock \bibinfo{title}{{T}axonomy --- openethics.ai}.
\newblock \bibinfo{howpublished}{\url{https://openethics.ai/taxonomy/}}.
\newblock
\newblock
\shownote{[Accessed 22-01-2024].}


\bibitem[\protect\citeauthoryear{Luo, Tong, Fang, and Qu}{Luo
  et~al\mbox{.}}{2019}]%
        {Luo2019}
\bibfield{author}{\bibinfo{person}{Xueming Luo}, \bibinfo{person}{Siliang
  Tong}, \bibinfo{person}{Zheng Fang}, {and} \bibinfo{person}{Zhe Qu}.}
  \bibinfo{year}{2019}\natexlab{}.
\newblock \showarticletitle{Frontiers: Machines vs. Humans: The Impact of
  Artificial Intelligence Chatbot Disclosure on Customer Purchases}.
\newblock \bibinfo{journal}{\emph{Marketing Science}} \bibinfo{volume}{38},
  \bibinfo{number}{6} (\bibinfo{year}{2019}), \bibinfo{pages}{937--947}.
\newblock
\urldef\tempurl%
\url{https://doi.org/10.1287/mksc.2019.1192}
\showDOI{\tempurl}
\showeprint{https://doi.org/10.1287/mksc.2019.1192}


\bibitem[\protect\citeauthoryear{McDonald, Schoenebeck, and Forte}{McDonald
  et~al\mbox{.}}{2019}]%
        {McDonald2019reliability}
\bibfield{author}{\bibinfo{person}{Nora McDonald}, \bibinfo{person}{Sarita
  Schoenebeck}, {and} \bibinfo{person}{Andrea Forte}.}
  \bibinfo{year}{2019}\natexlab{}.
\newblock \showarticletitle{Reliability and Inter-rater Reliability in
  Qualitative Research: Norms and Guidelines for CSCW and HCI Practice}.
\newblock \bibinfo{journal}{\emph{Proceedings of the ACM on Human-Computer
  Interaction}}  \bibinfo{volume}{3} (\bibinfo{date}{11} \bibinfo{year}{2019}),
  \bibinfo{pages}{1--23}.
\newblock
\urldef\tempurl%
\url{https://doi.org/10.1145/3359174}
\showDOI{\tempurl}


\bibitem[\protect\citeauthoryear{Monitor}{Monitor}{2024}]%
        {techmonitorTextLeaked}
\bibfield{author}{\bibinfo{person}{Tech Monitor}.}
  \bibinfo{year}{2024}\natexlab{}.
\newblock \bibinfo{title}{{T}ext of {E}{U} {A}{I} {A}ct leaked amid debate over
  the timeline for final approval --- techmonitor.ai}.
\newblock
  \bibinfo{howpublished}{\url{https://techmonitor.ai/technology/ai-and-automation/eu-ai-act-leaked-short-timeline}}.
\newblock
\newblock
\shownote{[Accessed 24-01-2024].}


\bibitem[\protect\citeauthoryear{Muralidhar, Belloum, de~Oliveira, and
  Ashok}{Muralidhar et~al\mbox{.}}{2023}]%
        {Muralidhar2023}
\bibfield{author}{\bibinfo{person}{Deepa Muralidhar}, \bibinfo{person}{Rafik
  Belloum}, \bibinfo{person}{Kathia~Mar{\c{c}}al de Oliveira}, {and}
  \bibinfo{person}{Ashwin Ashok}.} \bibinfo{year}{2023}\natexlab{}.
\newblock \showarticletitle{Elements that Influence Transparency
  in Artificial Intelligent Systems - A Survey}. In
  \bibinfo{booktitle}{\emph{Human-Computer Interaction -- INTERACT 2023}},
  \bibfield{editor}{\bibinfo{person}{Jos{\'e} Abdelnour~Nocera},
  \bibinfo{person}{Marta Krist{\'i}n~L{\'a}rusd{\'o}ttir},
  \bibinfo{person}{Helen Petrie}, \bibinfo{person}{Antonio Piccinno}, {and}
  \bibinfo{person}{Marco Winckler}} (Eds.). \bibinfo{publisher}{Springer Nature
  Switzerland}, \bibinfo{address}{Cham}, \bibinfo{pages}{349--358}.
\newblock
\showISBNx{978-3-031-42280-5}


\bibitem[\protect\citeauthoryear{Novelli, Casolari, Hacker, Spedicato, and
  Floridi}{Novelli et~al\mbox{.}}{2024}]%
        {Novelli2024generative}
\bibfield{author}{\bibinfo{person}{Claudio Novelli}, \bibinfo{person}{Federico
  Casolari}, \bibinfo{person}{Philipp Hacker}, \bibinfo{person}{Giorgio
  Spedicato}, {and} \bibinfo{person}{Luciano Floridi}.}
  \bibinfo{year}{2024}\natexlab{}.
\newblock \bibinfo{title}{Generative AI in EU Law: Liability, Privacy,
  Intellectual Property, and Cybersecurity}.
\newblock
\newblock
\showeprint[arxiv]{2401.07348}~[cs.CY]


\bibitem[\protect\citeauthoryear{of~the EU}{of~the EU}{2024}]%
        {AIActupdates}
\bibfield{author}{\bibinfo{person}{Council of~the EU}.}
  \bibinfo{year}{2024}\natexlab{}.
\newblock \bibinfo{title}{{A}rtificial intelligence act: {C}ouncil and
  {P}arliament strike a deal on the first rules for {A}{I} in the world}.
\newblock
  \bibinfo{howpublished}{\url{https://www.consilium.europa.eu/en/press/press-releases/2023/12/09/artificial-intelligence-act-council-and-parliament-strike-a-deal-on-the-first-worldwide-rules-for-ai/}}.
\newblock
\newblock
\shownote{[Accessed 22-01-2024].}


\bibitem[\protect\citeauthoryear{OpenAI}{OpenAI}{2023}]%
        {openai2023}
\bibfield{author}{\bibinfo{person}{OpenAI}.} \bibinfo{year}{2023}\natexlab{}.
\newblock \bibinfo{title}{GPT-4 Technical Report}.
\newblock
\newblock
\showeprint[arxiv]{2303.08774}~[cs.DL]


\bibitem[\protect\citeauthoryear{OpenAI}{OpenAI}{2024}]%
        {openaiOpenAIApproaching}
\bibfield{author}{\bibinfo{person}{OpenAI}.} \bibinfo{year}{2024}\natexlab{}.
\newblock \bibinfo{title}{{H}ow {O}pen{A}{I} is approaching 2024 worldwide
  elections --- openai.com}.
\newblock
  \bibinfo{howpublished}{\url{https://openai.com/blog/how-openai-is-approaching-2024-worldwide-elections}}.
\newblock
\newblock
\shownote{[Accessed 17-01-2024].}


\bibitem[\protect\citeauthoryear{Panigutti, Hamon, Hupont, Fernandez~Llorca,
  Fano~Yela, Junklewitz, Scalzo, Mazzini, Sanchez, Soler~Garrido, and
  Gomez}{Panigutti et~al\mbox{.}}{2023}]%
        {Panigutti2023}
\bibfield{author}{\bibinfo{person}{Cecilia Panigutti}, \bibinfo{person}{Ronan
  Hamon}, \bibinfo{person}{Isabelle Hupont}, \bibinfo{person}{David
  Fernandez~Llorca}, \bibinfo{person}{Delia Fano~Yela}, \bibinfo{person}{Henrik
  Junklewitz}, \bibinfo{person}{Salvatore Scalzo}, \bibinfo{person}{Gabriele
  Mazzini}, \bibinfo{person}{Ignacio Sanchez}, \bibinfo{person}{Josep
  Soler~Garrido}, {and} \bibinfo{person}{Emilia Gomez}.}
  \bibinfo{year}{2023}\natexlab{}.
\newblock \showarticletitle{The Role of Explainable AI in the Context of the AI
  Act}. In \bibinfo{booktitle}{\emph{Proceedings of the 2023 ACM Conference on
  Fairness, Accountability, and Transparency}} (Chicago, IL, USA)
  \emph{(\bibinfo{series}{FAccT '23})}. \bibinfo{publisher}{Association for
  Computing Machinery}, \bibinfo{address}{New York, NY, USA},
  \bibinfo{pages}{1139--1150}.
\newblock
\showISBNx{9798400701924}
\urldef\tempurl%
\url{https://doi.org/10.1145/3593013.3594069}
\showDOI{\tempurl}


\bibitem[\protect\citeauthoryear{Pennycook, Bear, Collins, and Rand}{Pennycook
  et~al\mbox{.}}{2020}]%
        {Pennycook2020}
\bibfield{author}{\bibinfo{person}{Gordon Pennycook}, \bibinfo{person}{Adam
  Bear}, \bibinfo{person}{Evan~T. Collins}, {and} \bibinfo{person}{David~G.
  Rand}.} \bibinfo{year}{2020}\natexlab{}.
\newblock \showarticletitle{The Implied Truth Effect: Attaching Warnings to a
  Subset of Fake News Headlines Increases Perceived Accuracy of Headlines
  Without Warnings}.
\newblock \bibinfo{journal}{\emph{Management Science}} \bibinfo{volume}{66},
  \bibinfo{number}{11} (\bibinfo{year}{2020}), \bibinfo{pages}{4944--4957}.
\newblock
\urldef\tempurl%
\url{https://doi.org/10.1287/mnsc.2019.3478}
\showDOI{\tempurl}
\showeprint{https://doi.org/10.1287/mnsc.2019.3478}


\bibitem[\protect\citeauthoryear{Rakova, Yang, Cramer, and Chowdhury}{Rakova
  et~al\mbox{.}}{2021}]%
        {Rakova2021responsibleai}
\bibfield{author}{\bibinfo{person}{Bogdana Rakova}, \bibinfo{person}{Jingying
  Yang}, \bibinfo{person}{Henriette Cramer}, {and} \bibinfo{person}{Rumman
  Chowdhury}.} \bibinfo{year}{2021}\natexlab{}.
\newblock \showarticletitle{Where Responsible AI Meets Reality: Practitioner
  Perspectives on Enablers for Shifting Organizational Practices}.
\newblock \bibinfo{journal}{\emph{Proc. ACM Hum.-Comput. Interact.}}
  \bibinfo{volume}{5}, \bibinfo{number}{CSCW1}, Article \bibinfo{articleno}{7}
  (\bibinfo{date}{apr} \bibinfo{year}{2021}), \bibinfo{numpages}{23}~pages.
\newblock
\urldef\tempurl%
\url{https://doi.org/10.1145/3449081}
\showDOI{\tempurl}


\bibitem[\protect\citeauthoryear{Sadasivan, Kumar, Balasubramanian, Wang, and
  Feizi}{Sadasivan et~al\mbox{.}}{2023}]%
        {sadasivan2023aigenerated}
\bibfield{author}{\bibinfo{person}{Vinu~Sankar Sadasivan},
  \bibinfo{person}{Aounon Kumar}, \bibinfo{person}{Sriram Balasubramanian},
  \bibinfo{person}{Wenxiao Wang}, {and} \bibinfo{person}{Soheil Feizi}.}
  \bibinfo{year}{2023}\natexlab{}.
\newblock \bibinfo{title}{Can AI-Generated Text be Reliably Detected?}
\newblock
\newblock
\showeprint[arxiv]{2303.11156}~[cs.CL]


\bibitem[\protect\citeauthoryear{Saharia, Chan, Saxena, Li, Whang, Denton,
  Ghasemipour, Gontijo~Lopes, Karagol~Ayan, Salimans, et~al\mbox{.}}{Saharia
  et~al\mbox{.}}{2022}]%
        {saharia2022photorealistic}
\bibfield{author}{\bibinfo{person}{Chitwan Saharia}, \bibinfo{person}{William
  Chan}, \bibinfo{person}{Saurabh Saxena}, \bibinfo{person}{Lala Li},
  \bibinfo{person}{Jay Whang}, \bibinfo{person}{Emily~L Denton},
  \bibinfo{person}{Kamyar Ghasemipour}, \bibinfo{person}{Raphael
  Gontijo~Lopes}, \bibinfo{person}{Burcu Karagol~Ayan}, \bibinfo{person}{Tim
  Salimans}, {et~al\mbox{.}}} \bibinfo{year}{2022}\natexlab{}.
\newblock \showarticletitle{Photorealistic text-to-image diffusion models with
  deep language understanding}.
\newblock \bibinfo{journal}{\emph{Advances in Neural Information Processing
  Systems}}  \bibinfo{volume}{35} (\bibinfo{year}{2022}),
  \bibinfo{pages}{36479--36494}.
\newblock


\bibitem[\protect\citeauthoryear{Scharowski, Benk, Kühne, Wettstein, and
  Brühlmann}{Scharowski et~al\mbox{.}}{2023}]%
        {scharowski_certification_2023}
\bibfield{author}{\bibinfo{person}{Nicolas Scharowski},
  \bibinfo{person}{Michaela Benk}, \bibinfo{person}{Swen~J. Kühne},
  \bibinfo{person}{Léane Wettstein}, {and} \bibinfo{person}{Florian
  Brühlmann}.} \bibinfo{year}{2023}\natexlab{}.
\newblock \showarticletitle{Certification {Labels} for {Trustworthy} {AI}:
  {Insights} {From} an {Empirical} {Mixed}-{Method} {Study}}. In
  \bibinfo{booktitle}{\emph{2023 {ACM} {Conference} on {Fairness},
  {Accountability}, and {Transparency}}}. \bibinfo{publisher}{ACM},
  \bibinfo{address}{Chicago IL USA}, \bibinfo{pages}{248--260}.
\newblock
\showISBNx{9798400701924}
\urldef\tempurl%
\url{https://doi.org/10.1145/3593013.3593994}
\showDOI{\tempurl}


\bibitem[\protect\citeauthoryear{Schmidt, Elagroudy, Draxler, Kreuter, and
  Welsch}{Schmidt et~al\mbox{.}}{2024}]%
        {Schmidt2024}
\bibfield{author}{\bibinfo{person}{Albrecht Schmidt}, \bibinfo{person}{Passant
  Elagroudy}, \bibinfo{person}{Fiona Draxler}, \bibinfo{person}{Frauke
  Kreuter}, {and} \bibinfo{person}{Robin Welsch}.}
  \bibinfo{year}{2024}\natexlab{}.
\newblock \showarticletitle{Simulating the Human in HCD with ChatGPT:
  Redesigning Interaction Design with AI}.
\newblock \bibinfo{journal}{\emph{Interactions}} \bibinfo{volume}{31},
  \bibinfo{number}{1} (\bibinfo{date}{jan} \bibinfo{year}{2024}),
  \bibinfo{pages}{24--31}.
\newblock
\showISSN{1072-5520}
\urldef\tempurl%
\url{https://doi.org/10.1145/3637436}
\showDOI{\tempurl}


\bibitem[\protect\citeauthoryear{Springer and Whittaker}{Springer and
  Whittaker}{2020}]%
        {springer_progressive_2020}
\bibfield{author}{\bibinfo{person}{Aaron Springer} {and} \bibinfo{person}{Steve
  Whittaker}.} \bibinfo{year}{2020}\natexlab{}.
\newblock \showarticletitle{Progressive {Disclosure}: {When}, {Why}, and {How}
  {Do} {Users} {Want} {Algorithmic} {Transparency} {Information}?}
\newblock \bibinfo{journal}{\emph{ACM Transactions on Interactive Intelligent
  Systems}} \bibinfo{volume}{10}, \bibinfo{number}{4} (\bibinfo{date}{Dec.}
  \bibinfo{year}{2020}), \bibinfo{pages}{1--32}.
\newblock
\showISSN{2160-6455, 2160-6463}
\urldef\tempurl%
\url{https://doi.org/10.1145/3374218}
\showDOI{\tempurl}


\bibitem[\protect\citeauthoryear{Sundar}{Sundar}{2020}]%
        {sundar2020rise}
\bibfield{author}{\bibinfo{person}{S~Shyam Sundar}.}
  \bibinfo{year}{2020}\natexlab{}.
\newblock \showarticletitle{Rise of machine agency: A framework for studying
  the psychology of human--AI interaction (HAII)}.
\newblock \bibinfo{journal}{\emph{Journal of Computer-Mediated Communication}}
  \bibinfo{volume}{25}, \bibinfo{number}{1} (\bibinfo{year}{2020}),
  \bibinfo{pages}{74--88}.
\newblock


\bibitem[\protect\citeauthoryear{Sundar and Kim}{Sundar and Kim}{2019}]%
        {Sundar2019}
\bibfield{author}{\bibinfo{person}{S.~Shyam Sundar} {and}
  \bibinfo{person}{Jinyoung Kim}.} \bibinfo{year}{2019}\natexlab{}.
\newblock \showarticletitle{Machine Heuristic: When We Trust Computers More
  than Humans with Our Personal Information}. In
  \bibinfo{booktitle}{\emph{Proceedings of the 2019 CHI Conference on Human
  Factors in Computing Systems}} (Glasgow, Scotland Uk)
  \emph{(\bibinfo{series}{CHI '19})}. \bibinfo{publisher}{Association for
  Computing Machinery}, \bibinfo{address}{New York, NY, USA},
  \bibinfo{pages}{1--9}.
\newblock
\showISBNx{9781450359702}
\urldef\tempurl%
\url{https://doi.org/10.1145/3290605.3300768}
\showDOI{\tempurl}


\bibitem[\protect\citeauthoryear{Toff and Simon}{Toff and Simon}{2023}]%
        {Toff2023-qq}
\bibfield{author}{\bibinfo{person}{Benjamin Toff} {and}
  \bibinfo{person}{Felix~M Simon}.} \bibinfo{year}{2023}\natexlab{}.
\newblock \bibinfo{title}{``Or they could just not use it?'': The Paradox of
  {AI} Disclosure for Audience Trust in News}.  (\bibinfo{date}{Dec.}
  \bibinfo{year}{2023}).
\newblock


\bibitem[\protect\citeauthoryear{Tohline et~al\mbox{.}}{Tohline
  et~al\mbox{.}}{2008}]%
        {tohline2008computational}
\bibfield{author}{\bibinfo{person}{Joel~E Tohline} {et~al\mbox{.}}}
  \bibinfo{year}{2008}\natexlab{}.
\newblock \showarticletitle{Computational provenance}.
\newblock \bibinfo{journal}{\emph{Computing in Science \& Engineering}}
  \bibinfo{volume}{10}, \bibinfo{number}{03} (\bibinfo{year}{2008}),
  \bibinfo{pages}{9--10}.
\newblock


\bibitem[\protect\citeauthoryear{Torres}{Torres}{2023}]%
        {Torres2023}
\bibfield{author}{\bibinfo{person}{Richie Torres}.}
  \bibinfo{year}{2023}\natexlab{}.
\newblock \bibinfo{title}{H.R.3831 - AI Disclosure Act of 2023}.
\newblock
\newblock
\urldef\tempurl%
\url{https://www.congress.gov/bill/118th-congress/house-bill/3831?s=1\&r=1}
\showURL{%
\tempurl}
\newblock
\shownote{Accessed: 2024-1-15.}


\bibitem[\protect\citeauthoryear{Umbrello and Poel}{Umbrello and Poel}{2021}]%
        {Umbrello2021valuedesign}
\bibfield{author}{\bibinfo{person}{Steven Umbrello} {and} \bibinfo{person}{Ibo
  Poel}.} \bibinfo{year}{2021}\natexlab{}.
\newblock \showarticletitle{Mapping value sensitive design onto AI for social
  good principles}.
\newblock \bibinfo{journal}{\emph{AI and Ethics}}  \bibinfo{volume}{1}
  (\bibinfo{date}{08} \bibinfo{year}{2021}), \bibinfo{pages}{3}.
\newblock
\urldef\tempurl%
\url{https://doi.org/10.1007/s43681-021-00038-3}
\showDOI{\tempurl}


\bibitem[\protect\citeauthoryear{Veale and Borgesius}{Veale and
  Borgesius}{2021}]%
        {VealeAIAct2021}
\bibfield{author}{\bibinfo{person}{Michael Veale} {and}
  \bibinfo{person}{Frederik~Zuiderveen Borgesius}.}
  \bibinfo{year}{2021}\natexlab{}.
\newblock \showarticletitle{Demystifying the Draft EU Artificial Intelligence
  Act: Analysing the good, the bad, and the unclear elements of the proposed
  approach}.
\newblock \bibinfo{journal}{\emph{Computer Law Review International}}
  \bibinfo{volume}{22}, \bibinfo{number}{4} (\bibinfo{year}{2021}),
  \bibinfo{pages}{97--112}.
\newblock
\urldef\tempurl%
\url{https://doi.org/10.9785/cri-2021-220402}
\showDOI{\tempurl}


\bibitem[\protect\citeauthoryear{Webster}{Webster}{2024}]%
        {DiscloseDefinition}
\bibfield{author}{\bibinfo{person}{Merriam Webster}.}
  \bibinfo{year}{2024}\natexlab{}.
\newblock \bibinfo{title}{Definition of {DISCLOSE}}.
\newblock
  \bibinfo{howpublished}{\url{https://www.merriam-webster.com/dictionary/disclose}}.
\newblock
\newblock
\shownote{Accessed: 2024-1-22.}


\bibitem[\protect\citeauthoryear{Yaqub, Kakhidze, Brockman, Memon, and
  Patil}{Yaqub et~al\mbox{.}}{2020}]%
        {Yaqub2020}
\bibfield{author}{\bibinfo{person}{Waheeb Yaqub}, \bibinfo{person}{Otari
  Kakhidze}, \bibinfo{person}{Morgan~L. Brockman}, \bibinfo{person}{Nasir
  Memon}, {and} \bibinfo{person}{Sameer Patil}.}
  \bibinfo{year}{2020}\natexlab{}.
\newblock \showarticletitle{Effects of Credibility Indicators on Social Media
  News Sharing Intent}. In \bibinfo{booktitle}{\emph{Proceedings of the 2020
  CHI Conference on Human Factors in Computing Systems}} (Honolulu, HI, USA)
  \emph{(\bibinfo{series}{CHI '20})}. \bibinfo{publisher}{Association for
  Computing Machinery}, \bibinfo{address}{New York, NY, USA},
  \bibinfo{pages}{1--14}.
\newblock
\showISBNx{9781450367080}
\urldef\tempurl%
\url{https://doi.org/10.1145/3313831.3376213}
\showDOI{\tempurl}


\bibitem[\protect\citeauthoryear{Zhong, Chang, Yang, Wu, Mahawaga~Arachchige,
  Pathmabandu, and Xue}{Zhong et~al\mbox{.}}{2023}]%
        {Zhong2023Copyright}
\bibfield{author}{\bibinfo{person}{Haonan Zhong}, \bibinfo{person}{Jiamin
  Chang}, \bibinfo{person}{Ziyue Yang}, \bibinfo{person}{Tingmin Wu},
  \bibinfo{person}{Pathum~Chamikara Mahawaga~Arachchige},
  \bibinfo{person}{Chehara Pathmabandu}, {and} \bibinfo{person}{Minhui Xue}.}
  \bibinfo{year}{2023}\natexlab{}.
\newblock \showarticletitle{Copyright Protection and Accountability of
  Generative AI: Attack, Watermarking and Attribution}. In
  \bibinfo{booktitle}{\emph{Companion Proceedings of the ACM Web Conference
  2023}} (Austin, TX, USA) \emph{(\bibinfo{series}{WWW '23 Companion})}.
  \bibinfo{publisher}{Association for Computing Machinery},
  \bibinfo{address}{New York, NY, USA}, \bibinfo{pages}{94–98}.
\newblock
\showISBNx{9781450394192}
\urldef\tempurl%
\url{https://doi.org/10.1145/3543873.3587321}
\showDOI{\tempurl}


\end{thebibliography}


\appendix
\onecolumn
\section{Appendix}

\begin{scriptsize}
\begin{longtable}{lp{4.5cm}lp{6cm}c}
\textbf{ID} & \textbf{Questions} & \textbf{Sub-theme} & \textbf{Theme} & \textbf{Votes} \\
\toprule
         1 & How much should you communicate? Does law provide any answers? If not what is ethical? & Ethical Implications of AI Use & Ethical, Legal, and Policy Considerations & 3 \\ 
        2 & How dangerous is the content generated by AI? & Ethical Implications of AI Use & Ethical, Legal, and Policy Considerations & 2 \\ 
        3 & Why should this be a right? & Ethical Implications of AI Use & Ethical, Legal, and Policy Considerations & 1 \\ 
        4 & Why is it important to disclose the limitation and complexity of AI systems? & Ethical Implications of AI Use & Ethical, Legal, and Policy Considerations &  \\ 
        5 & Whose interests have informed this law? & Ethical Implications of AI Use & Ethical, Legal, and Policy Considerations &  \\ 
        6 & Why should similar obligation not apply to human in digital setting? (think ``authenticity" in virtual worlds like 2nd life or WOW)" & Ethical Implications of AI Use & Ethical, Legal, and Policy Considerations &  \\ 
                        \cmidrule{1-5}
        7 & What should be disclosed? & Legal Compliance and AI Disclosure & Ethical, Legal, and Policy Considerations & 3 \\ 
        8 & Who is going to be responsible for the consequences? & Legal Compliance and AI Disclosure & Ethical, Legal, and Policy Considerations & 3 \\ 
        9 & Who is responsible for informing the user? & Legal Compliance and AI Disclosure & Ethical, Legal, and Policy Considerations & 3 \\ 
        10 & Who is going to compensate for damages? & Legal Compliance and AI Disclosure & Ethical, Legal, and Policy Considerations & 1 \\ 
        11 & Who decides what ``interaction" means? Art 52.1 what does it mean? & Legal Compliance and AI Disclosure & Ethical, Legal, and Policy Considerations & 1 \\ 
        12 & When should users be punished for not following the obligation? and when providers? & Legal Compliance and AI Disclosure & Ethical, Legal, and Policy Considerations & 1 \\ 
        13 & When to update the law given the fast technological development? & Legal Compliance and AI Disclosure & Ethical, Legal, and Policy Considerations & 1 \\ 
        14 & What are the intended uses and applications of AI systems as disclosed to users? & Legal Compliance and AI Disclosure & Ethical, Legal, and Policy Considerations &  \\ 
        15 & When is disclosure not sufficient? & Legal Compliance and AI Disclosure & Ethical, Legal, and Policy Considerations &  \\ 
        16 & When is communication authentic? & Legal Compliance and AI Disclosure & Ethical, Legal, and Policy Considerations &  \\ 
        17 & Why we need to make laws for disclosing of it? Why we let the provider and the user to determine by themselves? & Legal Compliance and AI Disclosure & Ethical, Legal, and Policy Considerations &  \\ 
        18 & Who should set the rule? & Legal Compliance and AI Disclosure & Ethical, Legal, and Policy Considerations &  \\ 
        19 & Who has to go to prison if something happens? & Legal Compliance and AI Disclosure & Ethical, Legal, and Policy Considerations &  \\ 
        20 & Who should contribute to develop the AI Disclosure? & Legal Compliance and AI Disclosure & Ethical, Legal, and Policy Considerations &  \\ 
        21 & Who is responsible for AI Disclosure? & Legal Compliance and AI Disclosure & Ethical, Legal, and Policy Considerations &  \\ 
        22 & Who should be disclosed? Context & Legal Compliance and AI Disclosure & Ethical, Legal, and Policy Considerations &  \\ 
                        \cmidrule{1-5}
        23 & Who is going to be impacted by non disclosure? & Policy and Regulatory Impact & Ethical, Legal, and Policy Considerations & 3 \\ 
        24 & When do users need to disclose the use of AI? Consider continued influence effect & Policy and Regulatory Impact & Ethical, Legal, and Policy Considerations & 2 \\ 
        25 & Who decides if a ``context of use" is obvious enough to not require an ``interaction"? & Policy and Regulatory Impact & Ethical, Legal, and Policy Considerations & 2 \\ 
        26 & What information should recipients not receive? & Policy and Regulatory Impact & Ethical, Legal, and Policy Considerations & 2 \\ 
        27 & When the AI disclosure should be introduced? & Policy and Regulatory Impact & Ethical, Legal, and Policy Considerations & 2 \\ 
        28 & What transparency level/degree needs to be provided to the users? & Policy and Regulatory Impact & Ethical, Legal, and Policy Considerations & 1 \\ 
        
                29 & Who is responsible for deciding if a disclosure is needed? & Policy and Regulatory Impact & Ethical, Legal, and Policy Considerations & 1 \\ 
        30 & When is it mandatory to disclose something is created by AI? Always? & Policy and Regulatory Impact & Ethical, Legal, and Policy Considerations & 1 \\ 
        31 & What is a good way to do AI disclosure in the context where the ``mystery?" is needed? & Policy and Regulatory Impact & Ethical, Legal, and Policy Considerations &  \\ 
        32 & What is the content that AI should disclose? & Policy and Regulatory Impact & Ethical, Legal, and Policy Considerations &  \\ 
        33 & Who wants to keep informed about the fact that is generated by AI? & Policy and Regulatory Impact & Ethical, Legal, and Policy Considerations &  \\ 
        34 & Who is going develop/fine tune the specific AI to create such contents? & Policy and Regulatory Impact & Ethical, Legal, and Policy Considerations &  \\ 
        35 & Where in the value chain should transparency be provided? & Policy and Regulatory Impact & Ethical, Legal, and Policy Considerations &  \\ 
        36 & Where is this monitored from? & Policy and Regulatory Impact & Ethical, Legal, and Policy Considerations &  \\ 
        37 & Why differentiating AI and Human is necessary? & Policy and Regulatory Impact & Ethical, Legal, and Policy Considerations &  \\ 
        38 & When to introduce AI disclosure for different AI systems? Such as google search, chatGPT? & Policy and Regulatory Impact & Ethical, Legal, and Policy Considerations &  \\ 
        39 & What timeline is needed to adopt the new ``disclosure" system? & Policy and Regulatory Impact & Ethical, Legal, and Policy Considerations &  \\ 
        40 & Who, within a chain of multiple actors, will be responsible for effectuating there disclosure obligations? & Policy and Regulatory Impact & Ethical, Legal, and Policy Considerations &  \\ 
        41 & When to provide AI disclosure for the users for different scenarios? & Policy and Regulatory Impact & Ethical, Legal, and Policy Considerations &  \\ 
        \midrule[1pt]
        42 & Who cares? & Evolving AI Technologies and Societal Impact & Future Considerations, Evolving Context, \& Practical Implementation & 7 \\ 
        43 & Who benefits from AI disclosure? & Evolving AI Technologies and Societal Impact & Future Considerations, Evolving Context, \& Practical Implementation & 2 \\ 
        44 & When could fake AI content actually affect users' opinion? & Evolving AI Technologies and Societal Impact & Future Considerations, Evolving Context, \& Practical Implementation &  \\ 
        45 & When can AI generated content affect the real life affairs? & Evolving AI Technologies and Societal Impact & Future Considerations, Evolving Context, \& Practical Implementation &  \\ 
        46 & Why did lawmakers opt for transparency/disclosures as the core solution to ``inauthentic" content? & Evolving AI Technologies and Societal Impact & Future Considerations, Evolving Context, \& Practical Implementation &  \\ 
        47 & When the AI disclosure agreement should work, avoid the effect of rapid development of AI technology? & Evolving AI Technologies and Societal Impact & Future Considerations, Evolving Context, \& Practical Implementation &  \\ 
        48 & Where could fake AI content show up? & Evolving AI Technologies and Societal Impact & Future Considerations, Evolving Context, \& Practical Implementation &  \\ 
                        \cmidrule{1-5}
        49 & Why is it important to inform people? & Future Trends and Legal Adaptation & Future Considerations, Evolving Context, \& Practical Implementation & 2 \\ 
        50 & Why should people care about this? & Future Trends and Legal Adaptation & Future Considerations, Evolving Context, \& Practical Implementation & 2 \\ 
        51 & Who is AI? & Future Trends and Legal Adaptation & Future Considerations, Evolving Context, \& Practical Implementation & 2 \\ 
        52 & What is the definition of AI generated content? (e.g. images, text) & Future Trends and Legal Adaptation & Future Considerations, Evolving Context, \& Practical Implementation &  \\ 
        53 & Why researchers are interested in AI systems disclosure? & Future Trends and Legal Adaptation & Future Considerations, Evolving Context, \& Practical Implementation &  \\ 
        54 & Why would AI have access to real world data? & Future Trends and Legal Adaptation & Future Considerations, Evolving Context, \& Practical Implementation &  \\ 
        55 & What training data is required for AI to create fake content? & Future Trends and Legal Adaptation & Future Considerations, Evolving Context, \& Practical Implementation &  \\ 
                        \cmidrule{1-5}
        56 & How much effort should be put into the classification? & Practical Challenges in AI Implementation & Future Considerations, Evolving Context, \& Practical Implementation & 2 \\ 
        57 & Where the AI disclosure is not welcomed? (e.g., Academia, Industry) & Practical Challenges in AI Implementation & Future Considerations, Evolving Context, \& Practical Implementation &  \\ 
        58 & What kind of AI model could be used in such situations? & Practical Challenges in AI Implementation & Future Considerations, Evolving Context, \& Practical Implementation &  \\ 
        59 & How to disclose implicitly or explicitly? & Practical Challenges in AI Implementation & Future Considerations, Evolving Context, \& Practical Implementation &  \\ 
        60 & How to correctly use AI for content production? & Practical Challenges in AI Implementation & Future Considerations, Evolving Context, \& Practical Implementation &  \\ 
        61 & How to trigger user feedback for AI disclosure or interactions & Practical Challenges in AI Implementation & Future Considerations, Evolving Context, \& Practical Implementation &  \\ 
           \midrule[1pt]
          62 & Why do providers need to disclose they use AI? & Ethical Considerations for Providers & Provider Responsibility and Industry Impact & 3 \\ 
        63 & Why wouldn't the providers want to disclose that it's an AI system? & Ethical Considerations for Providers & Provider Responsibility and Industry Impact & 2 \\ 
        64 & How should we treat users when not following the obligation? And how providers? (think punishment) & Ethical Considerations for Providers & Provider Responsibility and Industry Impact & 1 \\ 
        65 & Who determines the intention of the interaction? & Ethical Considerations for Providers & Provider Responsibility and Industry Impact &  \\ 
        66 & What type of content should not be labeled? & Ethical Considerations for Providers & Provider Responsibility and Industry Impact &  \\ 
        67 & Why provider should consider disclosure in their decision making process? & Ethical Considerations for Providers & Provider Responsibility and Industry Impact &  \\ 
                        \cmidrule{1-5}
        68 & Why do providers want to use AI? (what is the benefit?) & Industry-Specific Impact and Challenges & Provider Responsibility and Industry Impact & 3 \\ 
        69 & What is the effort required to disclose the information correctly? & Industry-Specific Impact and Challenges & Provider Responsibility and Industry Impact & 2 \\ 
        70 & When should providers start thinking about disclosure mechanisms in the production process? & Industry-Specific Impact and Challenges & Provider Responsibility and Industry Impact &  \\ 
        71 & What industry will be affected the most? & Industry-Specific Impact and Challenges & Provider Responsibility and Industry Impact &  \\ 
        72 & What is considered manipulation? Is there also a degree to this that is accounted for? & Industry-Specific Impact and Challenges & Provider Responsibility and Industry Impact &  \\ 
        73 & What exactly is a ``deep fake"? & Industry-Specific Impact and Challenges & Provider Responsibility and Industry Impact &  \\ 
                \cmidrule{1-5}
        74 & Where to label information / content? & Provider Disclosure Obligations & Provider Responsibility and Industry Impact & 1 \\ 
        75 & How can you know that the use of AI was not disclosed? & Provider Disclosure Obligations & Provider Responsibility and Industry Impact & 1 \\ 
        76 & When to not disclose information? & Provider Disclosure Obligations & Provider Responsibility and Industry Impact & 1 \\ 
        77 & What freedom shall the provider have in formulating/offering information to the recipient? & Provider Disclosure Obligations & Provider Responsibility and Industry Impact & 1 \\ 
        78 & What information should the provider disclose to the recipient of AI content & Provider Disclosure Obligations & Provider Responsibility and Industry Impact &  \\ 
        79 & Why would media/press announce that they are using AI? & Provider Disclosure Obligations & Provider Responsibility and Industry Impact &  \\ 
        80 & How to decide to what extent the provider does AI disclosure? & Provider Disclosure Obligations & Provider Responsibility and Industry Impact &  \\ 
        \midrule[1pt]
        81 & How important is it for users to know the authenticity of media? & Authenticity, Provenance, and Transparency & Trust, Authenticity, and User Empowerment & 2 \\ 
        82 & How are we going to verify authenticity? & Authenticity, Provenance, and Transparency & Trust, Authenticity, and User Empowerment & 2 \\ 
        83 & Who is going to verify the classification? & Authenticity, Provenance, and Transparency & Trust, Authenticity, and User Empowerment & 2 \\ 
        84 & Where can people manually verify if it is AI generated? & Authenticity, Provenance, and Transparency & Trust, Authenticity, and User Empowerment & 2 \\ 
        85 & What are the implications beyond the exact content? e.g job security for creatives? & Authenticity, Provenance, and Transparency & Trust, Authenticity, and User Empowerment & 1 \\ 
        86 & What types of risks can arise from transparency? (think privacy issues) & Authenticity, Provenance, and Transparency & Trust, Authenticity, and User Empowerment & 1 \\ 
        87 & What does ``authentic" content / information mean? & Authenticity, Provenance, and Transparency & Trust, Authenticity, and User Empowerment &  \\ 
        88 & Who might care the most about transparency? Users or providers and why? & Authenticity, Provenance, and Transparency & Trust, Authenticity, and User Empowerment &  \\ 
                \cmidrule{1-5}
        89 & How will different users interpret signals differently? & Building Trust in AI Systems & Trust, Authenticity, and User Empowerment & 1 \\ 
        90 & Where should users go if they have a complaint? & Building Trust in AI Systems & Trust, Authenticity, and User Empowerment &  \\ 
        91 & How does disclosure impact people's trust and perceived credibility? & Building Trust in AI Systems & Trust, Authenticity, and User Empowerment &  \\ 
        92 & How to responsibly disclose artistic uses of AI? & Building Trust in AI Systems & Trust, Authenticity, and User Empowerment &  \\ 
        93 & Why disclosure is considered as key part of user satisfaction? & Building Trust in AI Systems & Trust, Authenticity, and User Empowerment &  \\ 
        94 & What are the consequences of mis-disclosure? & Building Trust in AI Systems & Trust, Authenticity, and User Empowerment &  \\ 
        95 & How much does the user care? & Building Trust in AI Systems & Trust, Authenticity, and User Empowerment &  \\ 
        96 & Why would you choose this to communicate (any empirical studies to base your actions upon ?) & Building Trust in AI Systems & Trust, Authenticity, and User Empowerment &  \\ 
        97 & Why AI should be disclosed? & Building Trust in AI Systems & Trust, Authenticity, and User Empowerment &  \\ 
        98 & Why would people care about AI disclosure if AI content is flawless or productive? & Building Trust in AI Systems & Trust, Authenticity, and User Empowerment &  \\ 
                \cmidrule{1-5}
        99 & How can we train users to spot authentic content without relying on explicit cues? & Empowering Users through Education \& Awareness & Trust, Authenticity, and User Empowerment & 2 \\ 
        100 & What type of information empowers/triggers ``act of resistance"? & Empowering Users through Education \& Awareness & Trust, Authenticity, and User Empowerment & 2 \\ 
        101 & Why can't people distinguish themselves? & Empowering Users through Education \& Awareness & Trust, Authenticity, and User Empowerment & 2 \\ 
        102 & Where can people report false classification? & Empowering Users through Education \& Awareness & Trust, Authenticity, and User Empowerment & 1 \\ 
        103 & Where should we get ask if we suspect things (images, texts) are generated by AI? & Empowering Users through Education \& Awareness & Trust, Authenticity, and User Empowerment &  \\ 
        104 & How to balance AI generated ``True" and ``fake"? & Empowering Users through Education \& Awareness & Trust, Authenticity, and User Empowerment &  \\ 
        105 & What could be the most obvious feature for fake/AI generated content? & Empowering Users through Education \& Awareness & Trust, Authenticity, and User Empowerment &  \\ 
        106 & What are the features for fake/generated content? & Empowering Users through Education \& Awareness & Trust, Authenticity, and User Empowerment &  \\ 
        107 & What is the tolerated degree of user satisfaction for generated content? & Empowering Users through Education \& Awareness & Trust, Authenticity, and User Empowerment &  \\ 
        108 & Who could benefit from AI Disclosure? & Empowering Users through Education \& Awareness & Trust, Authenticity, and User Empowerment &  \\ 
        109 & Why people would care about AI disclosure or not? & Empowering Users through Education \& Awareness & Trust, Authenticity, and User Empowerment &  \\ 
                \cmidrule{1-5}
        110 & Why do users need to know they are interacting with AI? & User Responsibility and Agency & Trust, Authenticity, and User Empowerment & 3 \\ 
        111 & What information should the users acquire? Are they responsible for it? & User Responsibility and Agency & Trust, Authenticity, and User Empowerment & 1 \\ 
        112 & Why is it important for promoting democracy? & User Responsibility and Agency & Trust, Authenticity, and User Empowerment &  \\ 
        113 & Who is going to continue using AI to create Fake content? & User Responsibility and Agency & Trust, Authenticity, and User Empowerment &  \\ 
        114 & When to fully avoid the content without AI disclosure? & User Responsibility and Agency & Trust, Authenticity, and User Empowerment &  \\ 

   \midrule[1pt]
        
  115 & How much freedom should the users get to (not) see the classification? & Personalization and User Preferences & User Experience, Information Overload, and Personalization & 3 \\ 
        116 & How do users prefer to be informed? & Personalization and User Preferences & User Experience, Information Overload, and Personalization & 2 \\ 
        117 & How to adapt AI disclosures personally? & Personalization and User Preferences & User Experience, Information Overload, and Personalization & 1 \\ 
        118 & How to present them in different devices? & Personalization and User Preferences & User Experience, Information Overload, and Personalization &  \\ 
        119 & How to allow users to change the personalized UI settings? & Personalization and User Preferences & User Experience, Information Overload, and Personalization &  \\ 
        
\cmidrule{1-5}

        120 & Why do users trust AI more/less than a human? & Psychology of Human-AI Interaction & User Experience, Information Overload, and Personalization & 3 \\ 
        121 & How would the disclosure affect the users? & Psychology of Human-AI Interaction & User Experience, Information Overload, and Personalization & 1 \\ 
        122 & What is the psychological difference between AI and Human? & Psychology of Human-AI Interaction & User Experience, Information Overload, and Personalization &  \\ 
        123 & What are the differences users feel with fake or deep fake content? & Psychology of Human-AI Interaction & User Experience, Information Overload, and Personalization &  \\ 
        124 & Who are sensitive or insensitive for AI disclosure? & Psychology of Human-AI Interaction & User Experience, Information Overload, and Personalization &  \\ 
        125 & What is the criteria for the provider for balancing AI content and human content? & Psychology of Human-AI Interaction & User Experience, Information Overload, and Personalization &  \\ 
        \cmidrule{1-5}
        126 & What is a reliable method to test the classification? & Standardization & User Experience, Information Overload, and Personalization & 3 \\ 
        127 & Where on the website should Al be disclosed? Terms of service, every piece of content, every element? & Standardization & User Experience, Information Overload, and Personalization & 1 \\ 
        128 & Who determines how disclosure should be provided, according to which standards? & Standardization & User Experience, Information Overload, and Personalization & 1 \\ 
        129 & What would ensure that users are aware of the disclosure? (How will it be made? Another extensive terms and conditions or cookies? Do these currently work? & Standardization & User Experience, Information Overload, and Personalization &  \\ 
        130 & Who should assess if the AI disclosure is real/true? & Standardization & User Experience, Information Overload, and Personalization &  \\ 
        131 & How do you decide that the certification was designed by the right people and represents all relevant criteria? & Standardization & User Experience, Information Overload, and Personalization &  \\ 
        132 & Who communicates how the interaction and what stage of the interaction does this take place? & Standardization & User Experience, Information Overload, and Personalization &  \\ 
                \cmidrule{1-5}
        133 & Where is the classification going to be displayed? & User Interfaces and Information Overload & User Experience, Information Overload, and Personalization & 4 \\ 
        134 & Where is the balance between communicating enough and information overload? & User Interfaces and Information Overload & User Experience, Information Overload, and Personalization & 2 \\ 
        135 & How can you effectively communicate use of Al without distracting from content? & User Interfaces and Information Overload & User Experience, Information Overload, and Personalization & 1 \\ 
        136 & How can we make trust labels meaningful? & User Interfaces and Information Overload & User Experience, Information Overload, and Personalization & 1 \\ 
        137 & How to balance/control the information overload? & User Interfaces and Information Overload & User Experience, Information Overload, and Personalization &  \\ 
        138 & What would the effect of disclosure be? e.g., a page full of AI warnings may be off putting for the user & User Interfaces and Information Overload & User Experience, Information Overload, and Personalization &  \\ 
                \cmidrule{1-5}
        139 & Who are you communicating to? e.g., different people may require different types of communication mechanisms & User-Centric Information Design & User Experience, Information Overload, and Personalization & 4 \\ 
        140 & What under information is meaningful for the recipient? & User-Centric Information Design & User Experience, Information Overload, and Personalization & 2 \\ 
        141 & What format the users would like to see about the disclosure? & User-Centric Information Design & User Experience, Information Overload, and Personalization & 2 \\ 
        142 & Where to provide the AI disclosure? Before the generated content or after that? & User-Centric Information Design & User Experience, Information Overload, and Personalization & 2 \\ 
        143 & Where in people's living space should nature of whose clearly be communicated? (e.g, policy notice, screen)? & User-Centric Information Design & User Experience, Information Overload, and Personalization & 1 \\ 
        144 & How do you decide what type of information to communicate and who is your target group ? & User-Centric Information Design & User Experience, Information Overload, and Personalization & 1 \\ 
        145 & What is the balance between  AI generated ``fake" and ``true"? & User-Centric Information Design & User Experience, Information Overload, and Personalization &  \\ 
        146 & How to disclose it (AI/human made) by an implicit tag? & User-Centric Information Design & User Experience, Information Overload, and Personalization &  \\ 
        147 & How can we evaluate if the disclosure is sufficient enough information for the users? & User-Centric Information Design & User Experience, Information Overload, and Personalization &  \\ 
        148 & When do you engage in interdisciplinary collaborations between lawyers (so that you what and when to communicate) and those designing communications (so that lawyers know what's possible and propose policy changes) & User-Centric Information Design & User Experience, Information Overload, and Personalization &  \\ 
        149 & What kind of user feedback/ input is incorporated into the AI system as disclosed? & User-Centric Information Design & User Experience, Information Overload, and Personalization &  \\

\bottomrule
\\


\caption{Themes, sub-themes, and participant vote counts for the set of questions generated during our workshops.}
\Description{Themes, sub-themes, and participant vote counts for the set of 149 questions that were generated during our workshops.}
\label{table:questions1}

\end{longtable}

\end{scriptsize}

\clearpage
\twocolumn

\end{document}